\documentclass[twocolumn,tighten, table]{aastex631}
\usepackage{natbib}
\usepackage{amsmath}
\usepackage{listings}
\usepackage[normalem]{ulem}
\usepackage[section]{placeins}
\usepackage{color}
\usepackage{graphicx}
\usepackage{fancyvrb}
\usepackage{float}

\shorttitle{The dawn of disk formation}
\shortauthors{T.~Tamfal et al.}

\begin{document}

\title{\textbf{The Dawn of Disk Formation in a Milky Way-sized Galaxy Halo}: Thin Stellar Disks at $z > 4$}


\correspondingauthor{Tomas Tamfal}
\email{tomas.tamfal@uzh.ch}

\author[0000-0003-1773-9349]{Tomas Tamfal}
\affiliation{Center for Theoretical Astrophysics and Cosmology, Institute for Computational Science, University of Zurich, Winterthurerstrasse 190, CH-8057 Z\"urich, Switzerland}

\author[0000-0002-7078-2074]{Lucio Mayer}
\affiliation{Center for Theoretical Astrophysics and Cosmology, Institute for Computational Science, University of Zurich, Winterthurerstrasse 190, CH-8057 Z\"urich, Switzerland}

\author[0000-0001-5510-2803]{Thomas R. Quinn}
\affiliation{Astronomy Department, University of Washington, Seattle, WA 98195, USA}

\author[0000-0003-1746-9529]{Arif Babul}
\affiliation{Department of Physics \& Astronomy, University of Victoria, BC, V8X 4M6, Canada}

\author[0000-0002-6336-3293]{Piero Madau}
\affiliation{Department of Astronomy \& Astrophysics, University of California, 1156 High Street, Santa Cruz, CA 95064, USA}

\author[0000-0002-1786-963X]{Pedro R. Capelo}
\affiliation{Center for Theoretical Astrophysics and Cosmology, Institute for Computational Science, University of Zurich, Winterthurerstrasse 190, CH-8057 Z\"urich, Switzerland}

\author[0000-0001-8523-1171]{Sijing Shen}
\affiliation{Institute of Theoretical Astrophysics, University of Oslo, Postboks 1029, 0315 Oslo, Norway}

\author{Marius Staub}
\affiliation{Institute for Particle Physics and Astrophysics, Eidgen\"ossische Technische Hochschule, Wolfgang-Pauli-Strasse 27, 8049 Z\"urich, Switzerland}

\begin{abstract}
We present results from \textsc{GigaEris}, a cosmological, $N$-body hydrodynamical ``zoom-in'' simulation of the formation of a Milky Way-sized galaxy halo with unprecedented resolution, encompassing of order a billion particles within the refined region. The simulation employs a modern implementation of smoothed-particle hydrodynamics, including metal-line cooling and metal and thermal diffusion. We focus on the early assembly of the galaxy, down to redshift $z=4.4$. The simulated galaxy has properties consistent with extrapolations of the main sequence of star-forming galaxies to higher redshifts and levels off to a star formation rate of $\sim$60$\, M_{\odot}$~yr$^{-1}$ at $z=4.4$. A compact, thin rotating stellar disk with properties analogous to those of low-redshift systems arises already at $z \sim 8$. The galaxy rapidly develops a multi-component structure, and the disk, at least at these early stages, does not grow ``upside-down'' as often reported in the literature. Rather, at any given time, newly born stars contribute to sustain a thin disk. The kinematics reflect the early, ubiquitous presence of a thin disk, as a stellar disk component with $v_\phi/\sigma_R$ larger than unity is already present at $z \sim 9$--10. Our results suggest that high-resolution spectro-photometric observations of very high-redshift galaxies should find thin rotating disks, consistent with the recent discovery of cold rotating gas disks by ALMA. Finally, we present synthetic images for the JWST NIRCam camera, showing how the early disk would be easily detectable already at those early times.
\end{abstract}


\section{Introduction}\label{sec:Introduction}

\begin{figure*}
\includegraphics[width=\textwidth]{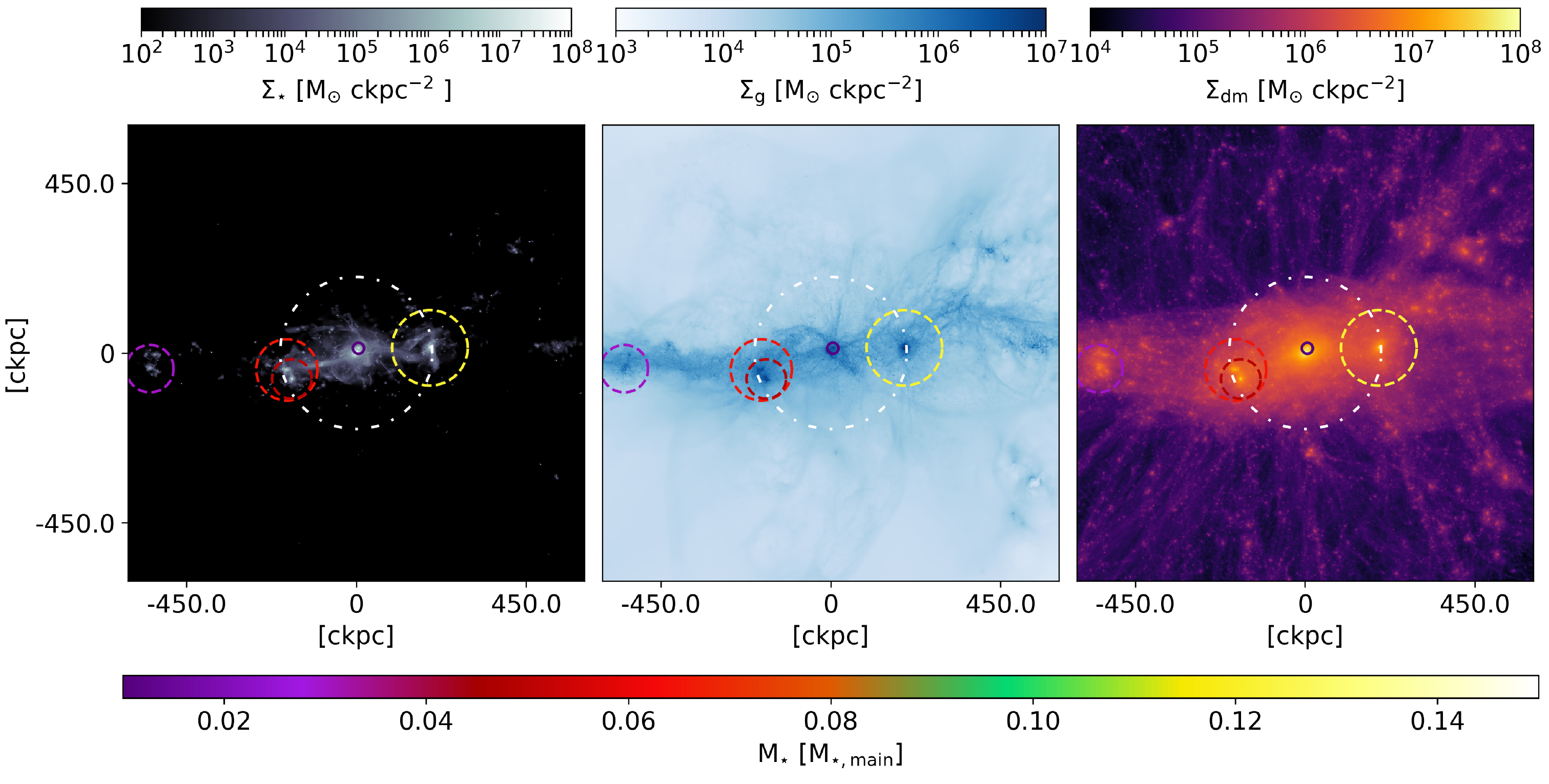}
\caption{The main halo of the \textsc{GigaEris} simulation at $z=4.44$. The left-hand panel shows the stellar surface density, the middle panel shows the gas surface density, and the right-hand panel depicts the DM surface density, all in units of M$_\odot$ ckpc$^{-2}$ (see \textit{top} colorbars). The white, dash-dotted circle depicts the virial radius of the main galaxy. The other circles depict the virial radii of subhalos and are solid if the entire subhalo is within the main halo's virial radius or dashed if it is partially outside. The colorbar of the circles (see \textit{bottom} colorbar) indicates the mass ratio between the stellar mass of the main halo and that of the subhalo.}
\label{fig:total}
\end{figure*}

The origin of the thin and thick disks of present-day spirals, most notably the Milky Way, has been subject of intense debate in the last several decades. Various mechanisms have been proposed to produce a thick disk out of the thin disk, which have been tested primarily via observations of the Milky Way, including heating by merging satellites, internal dynamical heating by buckling bars, and accretion of lower angular momentum stars from satellites themselves \citep[see][]{Gilmore_et_al_1989, Quinn_et_al_1993,Wyse_2001, Abadi_et_al_2003, Meza_et_al_2005,Wyse_et_al_2006,Villalobos_et_al_2008,Purcell_et_al_2009,DiMatteo_et_al_2011}. These mechanisms all play a role to some extent. On the numerical simulations side, in the last decade cosmological hydrodynamical simulations using the ``zoom-in'' technique have reached enough realism to be able to capture quite faithfully all these processes, as well as the re-formation of thin disk components from subsequent gas accretion episodes, especially at lower redshift, when gas accretes on higher angular momentum orbits \citep[][]{Bird_et_al_2013, Bird_et_al_2021, Grand_et_al_2016, Wetzel_et_al_2016, Hopkins_et_al_2018, El_Badry_et_al_2018, Hafen_et_al_2019}. Overall, simulations seem to have no more issues in producing a variety of disk galaxies whose disk structure resembles real ones \citep[][]{Grand_et_al_2016, Sokolowska_et_al_2017}. Such simulations also predict that, at high redshift, the early disk component is thick as the gas accretes in a rather turbulent environment with many mergers occurring in relatively short timescales and along different planes \citep[][]{Bird_et_al_2013}.

This has led to the notion that the disk forms \textit{``upside-down''}, namely the extended thin disk forms only at a later stage, when gas accretes along smoother, higher angular momentum filaments. Recently, however, observations with ALMA and other instruments have revealed the existence of rotationally supported disks of gas at very high redshift, bringing up the question of whether this is consistent or not with the \textit{``upside-down''} scenario \citep[e.g.][]{Hodge_et_al_2019, Rizzo_et_al_2020, Neeleman_et_al_2020,LeFevre_et_al_2020}. And indeed, lately new studies have shown that such a (gaseous) disk can also be found in numerical simulations \citep[see][]{Meng_et_al_2019, Meng_et_al_2021, Kretschmer_et_al_2021}. The importance of both mass and spatial resolution in understanding galaxy formation at low redshift, and in particular the assembly of the galactic disk component, has been shown clearly by the way progress has occurred throughout the 2000s \citep[e.g.][]{Kaufmann_et_al_2007, Mayer_et_al_2008, Governato_et_al_2010}. Similar issues might now be relevant to understand galaxy formation at much higher redshift.

Here, we revisit the problem of early disk assembly with a ``zoom-in'' cosmological hydrodynamical simulation of unprecedented resolution, \textsc{GigaEris}. Compared to predecessor runs such as those of the \textsc{Eris} suite, which were originally used to develop the scenario of \textit{``upside-down''} disk formation \citep[][]{Bird_et_al_2013}, \textsc{GigaEris} has a mass resolution nearly 20 times better, yielding a typical hydrodynamical spatial resolution of a few pc as opposed to tens of pc in previous simulations. Compared to the most recent ``zoom-in'' simulations of Milky Way-sized galaxy halos in the \textsc{FIRE-2} and \textsc{ELVIS-on-FIRE} suites \citep[][]{Garrison_et_al_2014, Garrison_et_al_2019, Santistevan_et_al_2021}, the gas mass resolution in \textsc{GigaEris} is, respectively, 8 times and 4 times better. This allows us to probe the early stages of galaxy formation with much greater physical realism than before, when the progenitors of present-day Milky Way-like galaxies have a baryonic and dark matter (DM) mass considerably smaller than at the present day. Equipped with such resolution, we can address whether the absence of a thin disk at high redshift in previous simulated galaxies is a physical result or the consequence of numerical limitations during the early stages of galaxy assembly which degrade the resolution of the scale height of the cold interstellar medium.

The paper is divided in four parts which are organized as follows; methods, results, discussion, and a summary section. Furthermore, the results section is again divided in three parts: a subsection about the evolution of the main galaxy as well as a comparison with previous work and observations, a subsection about the thin disk, and a subsection on the multiple formation mechanisms at high redshift.

\section{Methods}\label{s:plots}

\begin{figure}
\hspace{-1.5mm}\includegraphics[width=0.455\textwidth]{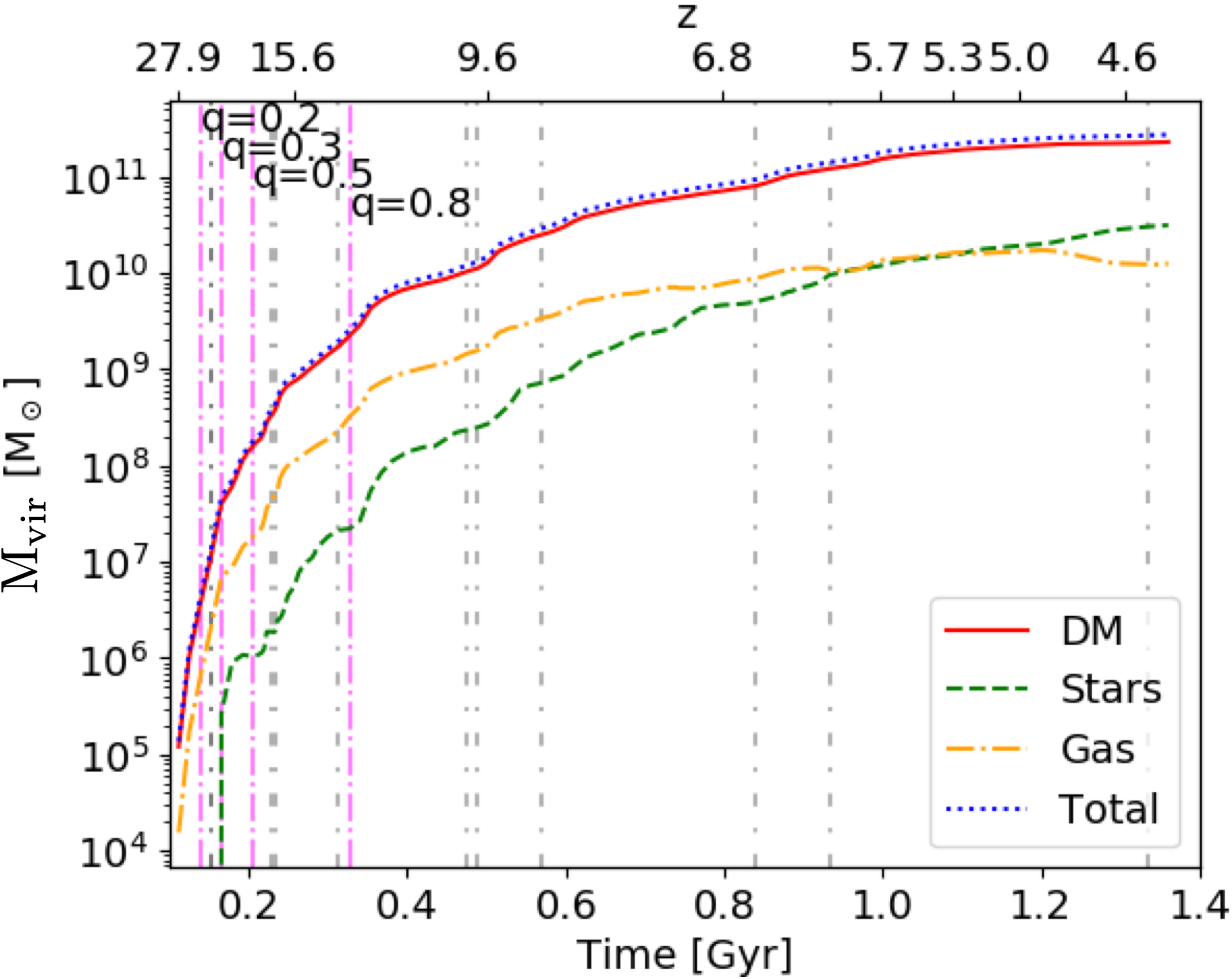}\\
\includegraphics[width=0.45\textwidth]{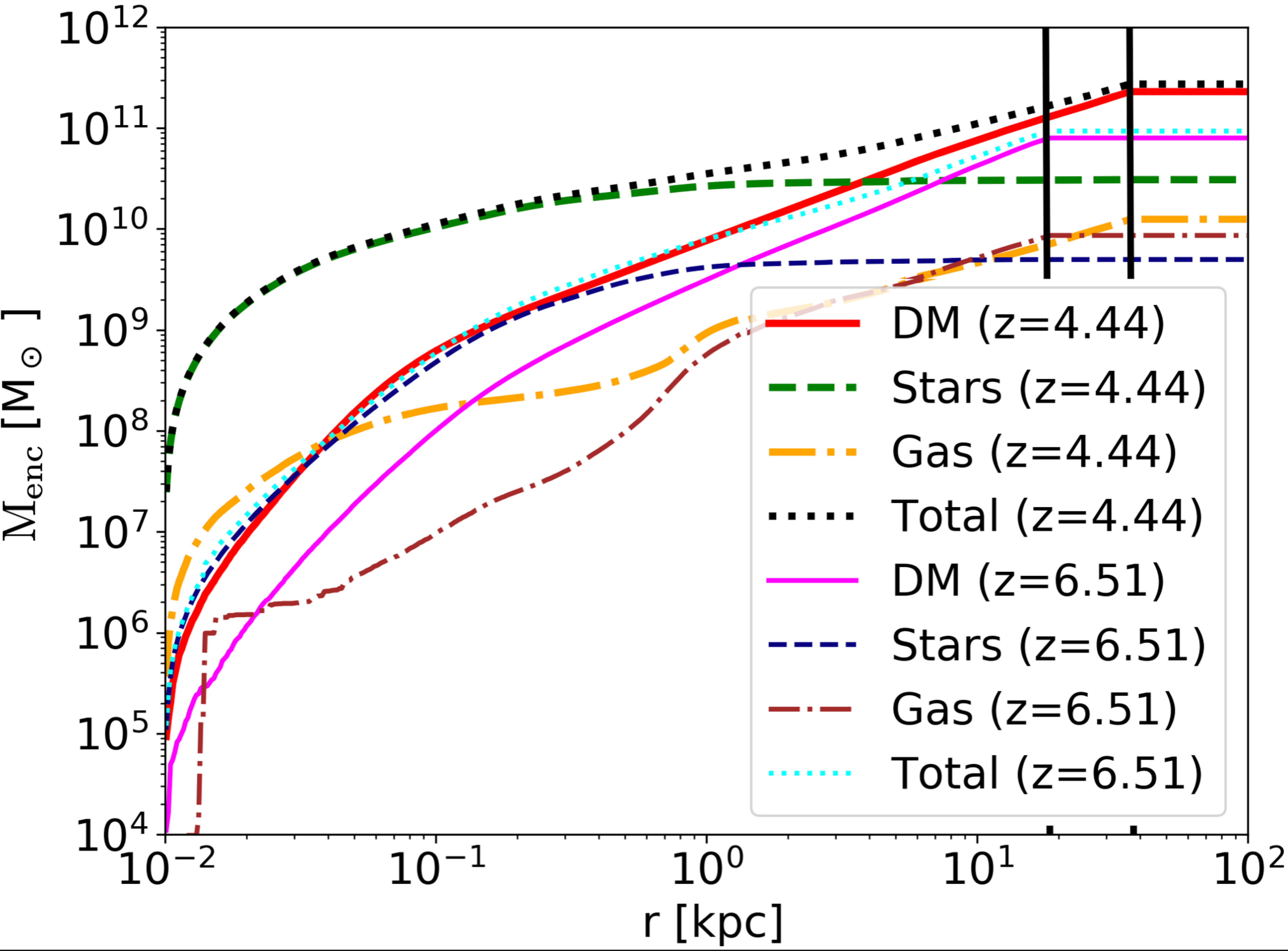}
\caption{The \textit{bottom} panel of this Figure depicts the enclosed mass of individual components as a function of radius at the end of our simulation ($z=4.44$) and at $z=6.51$: total mass (\textit{black} and \textit{cyan}, respectively), DM mass (\textit{red} and \textit{magenta}), stellar mass (\textit{green} and \textit{navy}), as well as gas mass (\textit{orange} and \textit{brown}). The two solid black vertical lines show the virial radius of the halo at the two selected redshifts. The \textit{top} panel shows the enclosed mass, within the virial radius, of individual components as a function of time using the same color coding. The dashed vertical lines depict mergers with a mass ratio $q \equiv \text{M}_\text{sat}/\text{M}_\text{halo} >0.1$ and the magenta lines highlight mergers with $q>0.25$.}
\label{fig:mass}
\end{figure}

\subsection{Simulation code}
\begin{figure}
\centering
\includegraphics[width=0.48\textwidth]{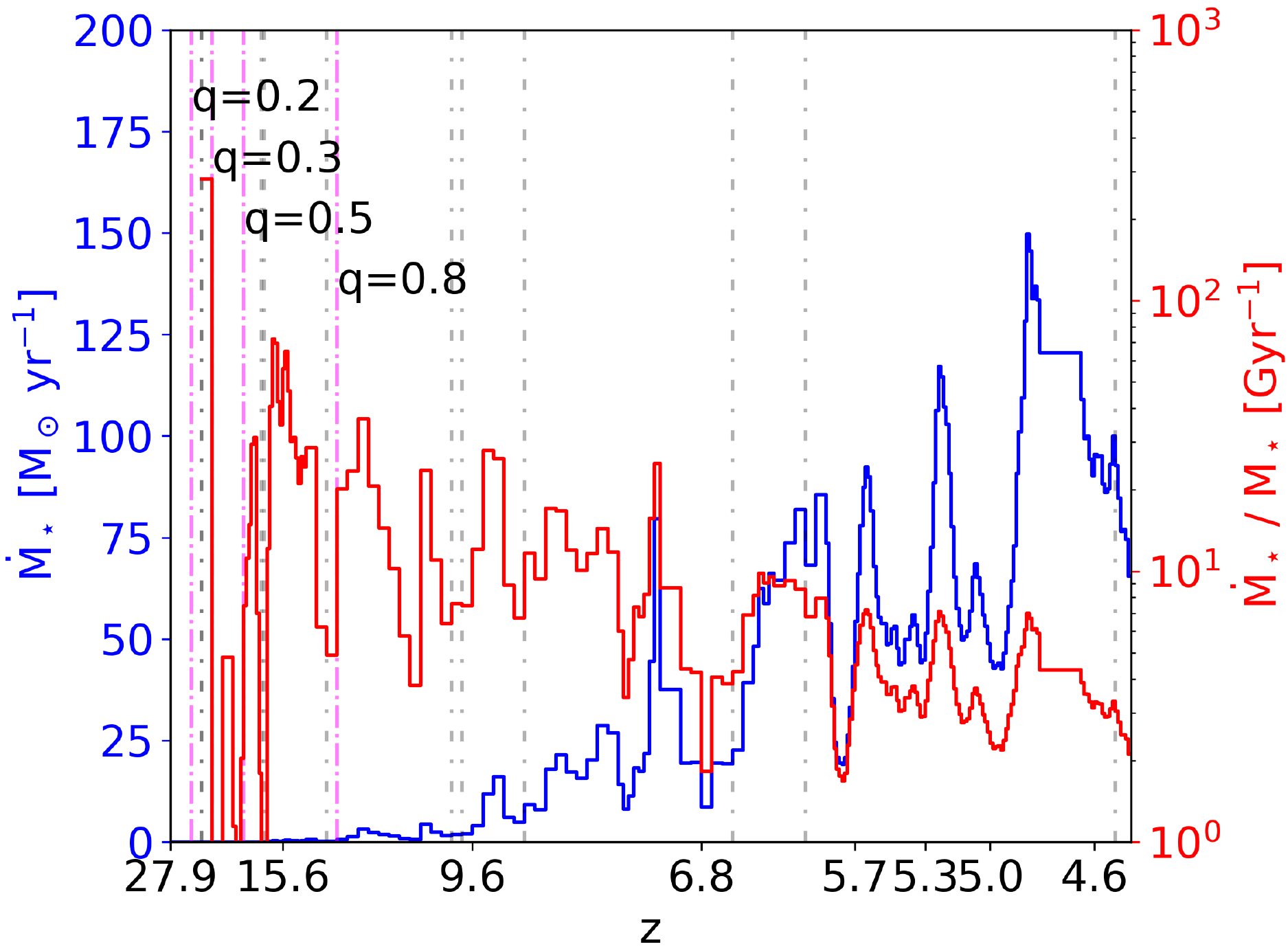}\\
\includegraphics[width=0.5\textwidth]{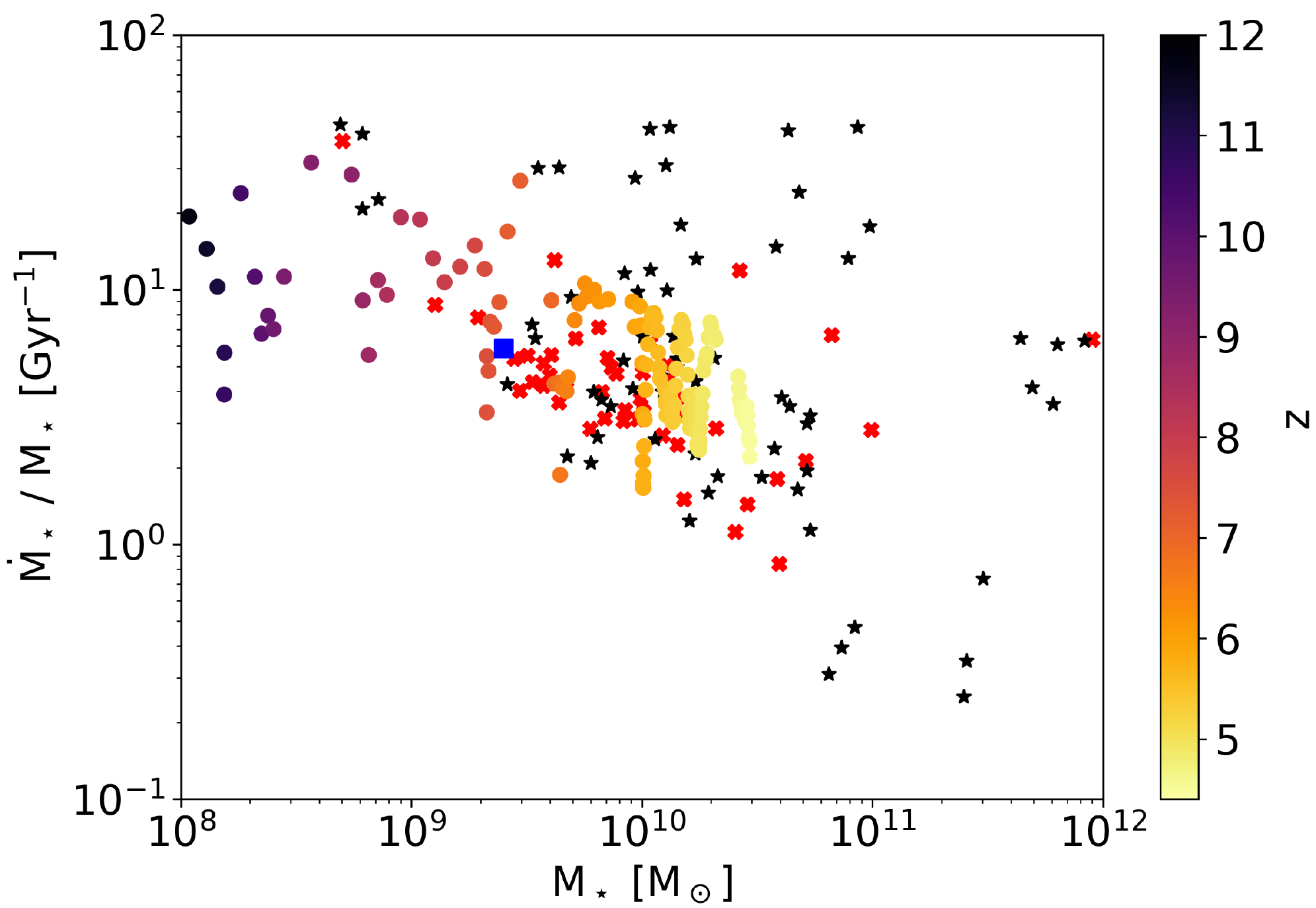}\\
\caption{\textit{Top panel:} Star formation rate (\textit{left-hand $y$-axis, blue curve}) and specific star formation rate (\textit{right-hand y-axis, red curve}) of the stars within a [6 kpc x 6 kpc x 4 kpc] central box of the main galaxy as a function of redshift. The vertical lines depict major mergers ($q>0.1$). \textit{Bottom panel:} Specific star formation rate (of the same subset of the top panel) as a function of stellar mass and color-coded with halo redshift. The VUDS survey data for galaxies in the redshift range $4.5 < z < 5.5$ \citep[see][]{Tasca_et_al_2015} are depicted as black stars (with a redshift determination that is $\approx 70$--75\% reliable) and red crosses (redshift determination is $\approx$~100\% reliable). This figure shows also the result of the \textsc{PONOS-HR} simulation at redshift $z=6.5$ as a blue square \citep[from][]{Fiacconi_et_al_2017}.}
\label{fig:SFH}
\end{figure}

The simulation was evolved with the collisionless $N$-body smoothed-particle hydrodynamics (SPH) code \textsc{ChaNGa} \citep[Charm $N$-body GrAvity solver, see][]{Jetley1, Jetley2, Menon_et_al_2015}. The code uses a \citet{BarnesHut1986} oct-tree to calculate gravity, with hexadecapole expansion of nodes and \citet{Ewald1921} summation for periodic forces. It follows the gas dynamics using a modern implementation of the SPH method, adopting a Wendland C4 kernel \citep[][]{Wendland1995,DehnenAly2012,Keller2014}, a geometric density average force calculation, and turbulent diffusion of thermal energy, as well as metal diffusion \citep{Wadsley_et_al_2017}. The timestepping in \textsc{ChaNGa} is carried out with a leapfrog integrator using individual timesteps for every particle. Each individual stellar particle represents an entire stellar population following the initial mass function (IMF) described in \citet{Kroupa_2001}, with an initial particle mass of $m_\star =1026$~M$_\odot$. We form stars stochastically using a simple gas density and temperature threshold criterion, with $n_{\rm SF} > 100$ atoms cm$^{-3}$ and $T < 3 \times 10^4$~K, and with a star formation rate given by

\begin{equation}\label{eq:schmidt}
\frac{\text{d} \rho_\star}{ \text{d} t }= \epsilon_{\rm SF} \frac{\rho_{\rm gas}}{t_{\rm dyn}},
\end{equation}

\noindent with $\rho_\star$ denoting the stellar density, $\rho_{\rm gas}$ the gas density, $t_{\rm dyn}$ the local dynamical time, and $\epsilon_{\rm SF} = 0.1$ the star formation efficiency. We also apply self-shielding \citep[see][]{Pontzen_et_al_2008} and a redshift-dependent ultra-violet (UV) radiation background \citep[][]{Haardt_et_al_2012}. In \textsc{GigaEris}, we solve for the non-equilibrium abundances of H and He ions, while cooling from fine-structure lines of metals at all temperatures (between $100$ and $10^{9}$~K) is calculated in photoionization equilibrium using tabulated rates from the \textsc{Cloudy} \citep[see][]{Ferland_et_al_1998, Ferland_et_al_2013}, following the method described in \cite{Shen_et_al_2010,Shen2013}. Feedback from supernovae Type Ia is implemented via stellar winds, whereas the feedback from supernovae Type II is implemented following the delayed-cooling recipes of \citet{Stinson_et_al_2006}, and injecting metals and $\epsilon_{\rm SN} = 10^{51}$~erg per event into the interstellar medium. The simulation used nearly 50 million core hours on the hybrid CPU+GPU nodes of the supercomputer Piz Daint at the Swiss National Supercomputing Centre (CSCS) and reached a final redshift of $z = 4.44$.

\subsection{Initial conditions}

In this work, we follow a galactic-scale halo identified at $z=0$ in a periodic cube of side 90~cMpc. We re-simulate the selected halo at several orders of magnitude higher resolution than the DM-only simulation, adding gas particles as well as the necessary short wavelength modes, by tracing back the  particles in an appropriately defined Lagrangian volume surrounding it, from $z=300$ until redshift $z=4.44$ ($\approx 1.4$~Gyr after the Big Bang). The galaxy halo was selected in a similar way as in the original \textsc{Eris} suite \citep[][]{Guedes_et_al_2011, Sokolowska_et_al_2017}, namely as one with a quiet merging history (last major merger at $z \sim 2$) and average spin parameter (0.045), with a halo virial mass similar but slightly larger than that of \textsc{Eris} at $z = 0$, $1.4 \times 10^{12} M_{\odot}$. At the early epochs that are the focus of this paper, the galaxy is located in one major filament (Figure~\ref{fig:total}). The initial conditions were created with the \textsc{MUSIC} code \citep[see][]{Hahn_et_al_2011}, using 14 levels of refinement and the cosmological parameters $\Omega_{\rm m} = 0.3089$, $\Omega_{\rm b} = 0.0486$, $\Omega_\Lambda = 0.6911$, $\sigma_8 = 0.8159$, $n_{\rm s} = 0.9667$, and $H_0 = 67.74$~km s$^{-1}$ Mpc$^{-1}$ \citep[see][]{Planck_et_al_2016}. The Lagrangian region for the high-resolution region was chosen to be an ellipsoid with a semi-minor axis of approximately 1.8 cMpc and a semi-major axis of 5~cMpc. The gravitational softening was chosen to be constant in physical coordinates for redshifts smaller than $z=10$, $\epsilon_{\rm C} = 0.043$~kpc, and otherwise to evolve as $\epsilon = 11 \epsilon_{\rm C} /(1+z)$. For the final snapshot at $z=4.44$, the particle numbers in the entire simulation are $n_{\rm DM} = 5.7 \times 10^{8}$, $n_{\rm gas} = 5.2 \times 10^{8}$, and $n_{\star} = 4.4 \times 10^{7}$, which leads to more than a billion ($1.1 \times 10^{9}$) particles. At the same time, within the virial radius, we obtain $m_{\rm DM} = 5493$~M$_\odot$ and $n_{\rm DM} = 4.2 \times 10^{7}$ for DM, a mean gas mass of $m_{\rm gas} = 1099$~M$_\odot$ and $n_{\rm gas} = 1.1 \times 10^{7}$, and a mean stellar mass of $m_\star = 798$~M$_\odot$ and $n_{\star} = 3.9 \times 10^{7}$. Note that with our simulation design the refined simulation volume is large enough to allow following the galaxy till $z=0$ without contamination by low-resolution particles.

\subsection{Halo Finding}

In order to find the main halo and also other substructures, we used the \textsc{AMIGA} Halo Finder \citep[hereafter AHF;][]{Gill_et_al_2004, Knollmann_et_al_2009} and applied it to our simulation box  with a minimum of 1000 DM particles, 10 star particles, and 10 gas particles per halo. 

\section{Results}
\begin{figure*}[htp]
\includegraphics[width=1\textwidth]{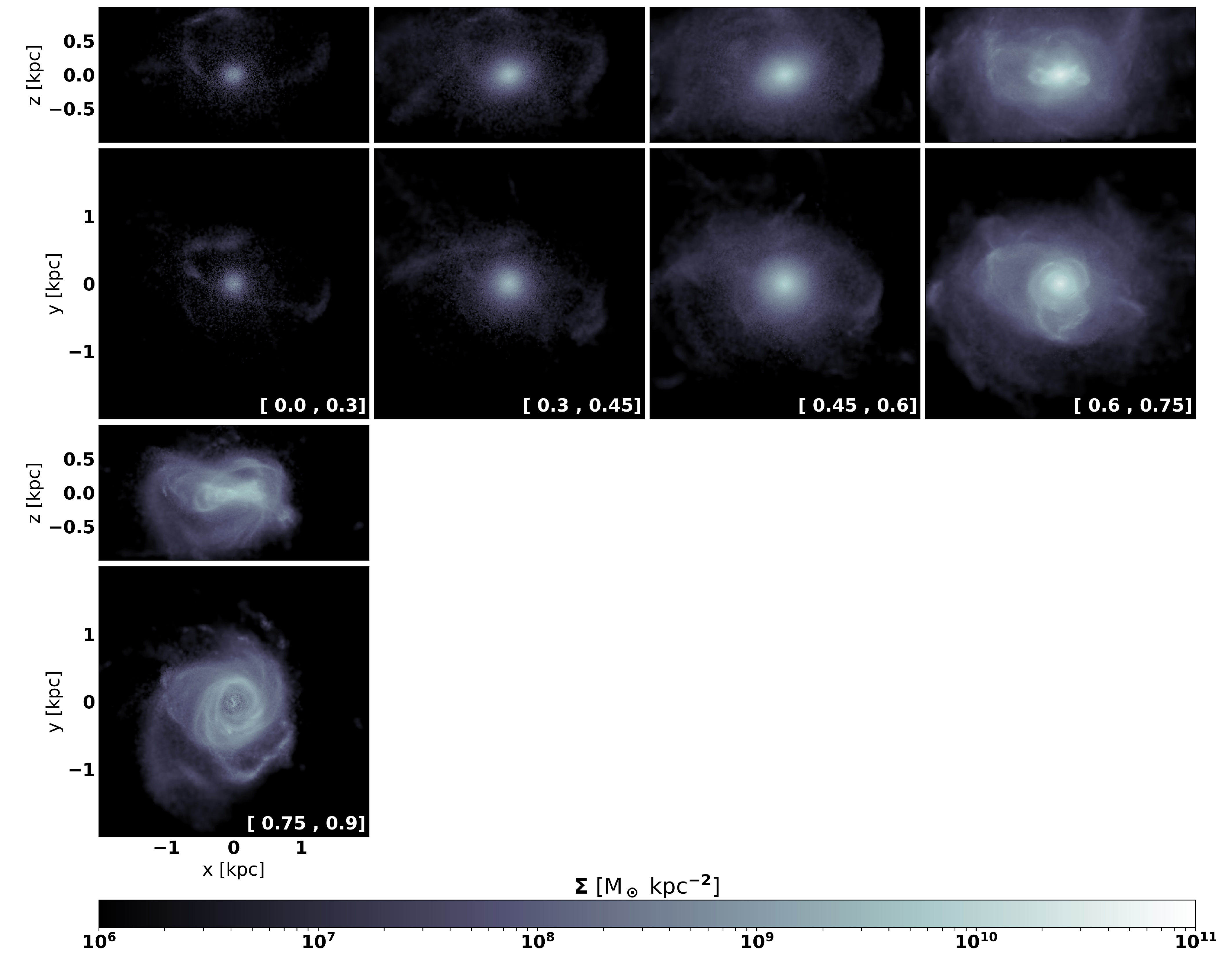}
\caption{Stellar surface density maps at $z=6.95$, approximately 0.77~Gyr after the Big Bang. The maps are oriented ``edge-on'' (\textit{top panels}) and ``face-on'' (\textit{bottom panels}), with the total stellar angular momentum along the $z$-axis, and afterwards split in various formation time bins (in Gyr). Each row shows surface densities in different bins associated to varying star formation time intervals (in Gyr), ranging from 0--0.3~Gyr (\textit{top left}) to 0.75--0.9~Gyr after the Big Bang (\textit{bottom left}). We should note that the disk, as can be inspected by the edge-on surface density maps, has a relatively large aspect ratio, which is at variance with present-day thin disks. However, the vertical extent of the disk, in the range 500--600~pc, does correspond to a present-day thin disk \citep[e.g.][]{Juric_et_al_2008}, which is why here, and throughout the text, we refer to it as a thin disk.}
\label{fig:ALL_Bird2}
\end{figure*}

\begin{figure*}[htp]
\includegraphics[width=1\textwidth]{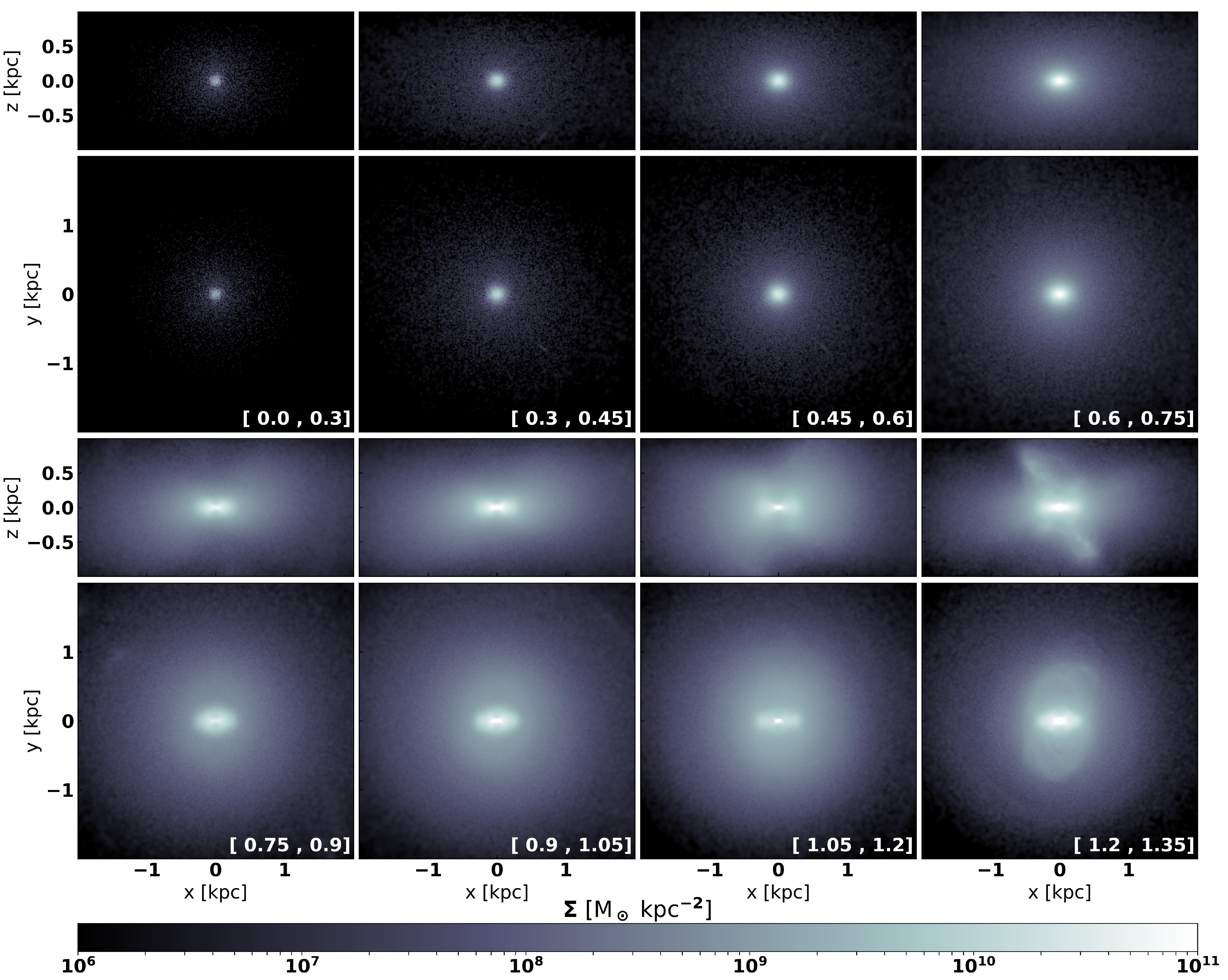}
\caption{Same as Figure~\ref{fig:ALL_Bird2}, but at $z =4.44$; the maps are oriented ``edge-on'' (\textit{top panels}) and ``face-on'' (\textit{bottom panels}), with the total stellar angular momentum along the $z$-axis, and afterwards split in various formation time bins (in Gyr). In contrast to Figure~\ref{fig:ALL_Bird2}, this Figure is created at the final snapshot of the simulation and therefore the associated star formation time intervals (in Gyr) are now ranging from 0--0.3~Gyr (\textit{top left}) to 1.2--1.35~Gyr after the Big Bang (\textit{bottom right}). Choosing a later time for this plot changes the interpretation, in contrast to Figure~\ref{fig:ALL_Bird2}, strikingly; we now do not obtain a thin disk forming (lower two rows of the plot) but a rather thick structure which seems to confirm the picture given in \cite{Bird_et_al_2013}. This can be attributed to the fact that we now have additional stars in the disk that could have been born ex-situ and therefore also change the visual appearance of each time bin.}
\label{fig:ALL_Bird}
\end{figure*}

\begin{figure}[htp]
\centering
\includegraphics[width=0.45\textwidth]{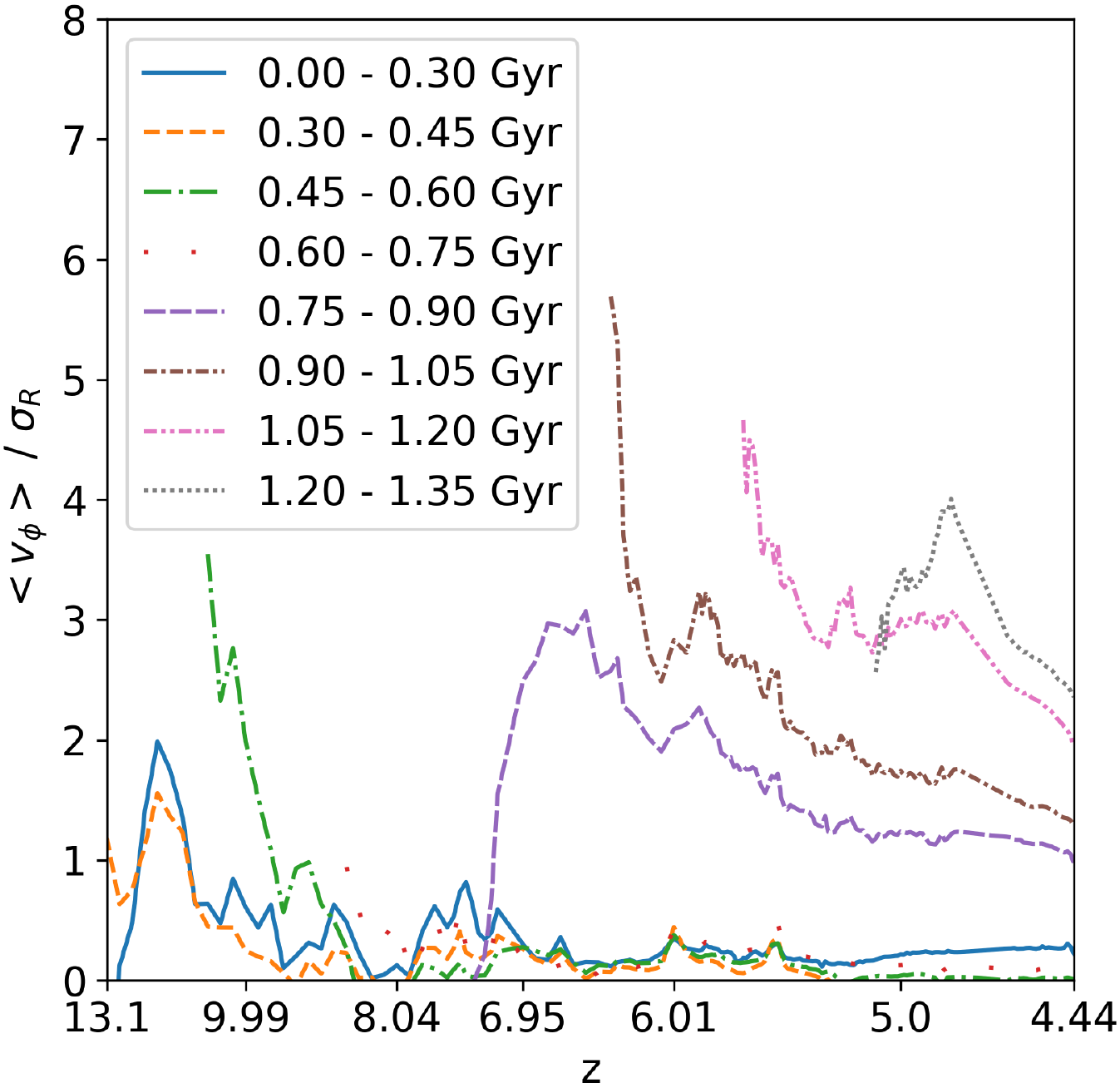}
\caption{The mean circular velocity $<v_\phi>$ divided by the radial velocity dispersion $\sigma_{R}$, binned by different stellar formation times as a function of time. The binning of the formation time is chosen to be the same as in Figure~\ref{fig:ALL_Bird}. After 0.8~Gyr, stars are always born in a disk-like structure.}
\label{fig:vcirc}
\end{figure}

\begin{figure}[htp]
\centering
\includegraphics[width=0.45\textwidth]{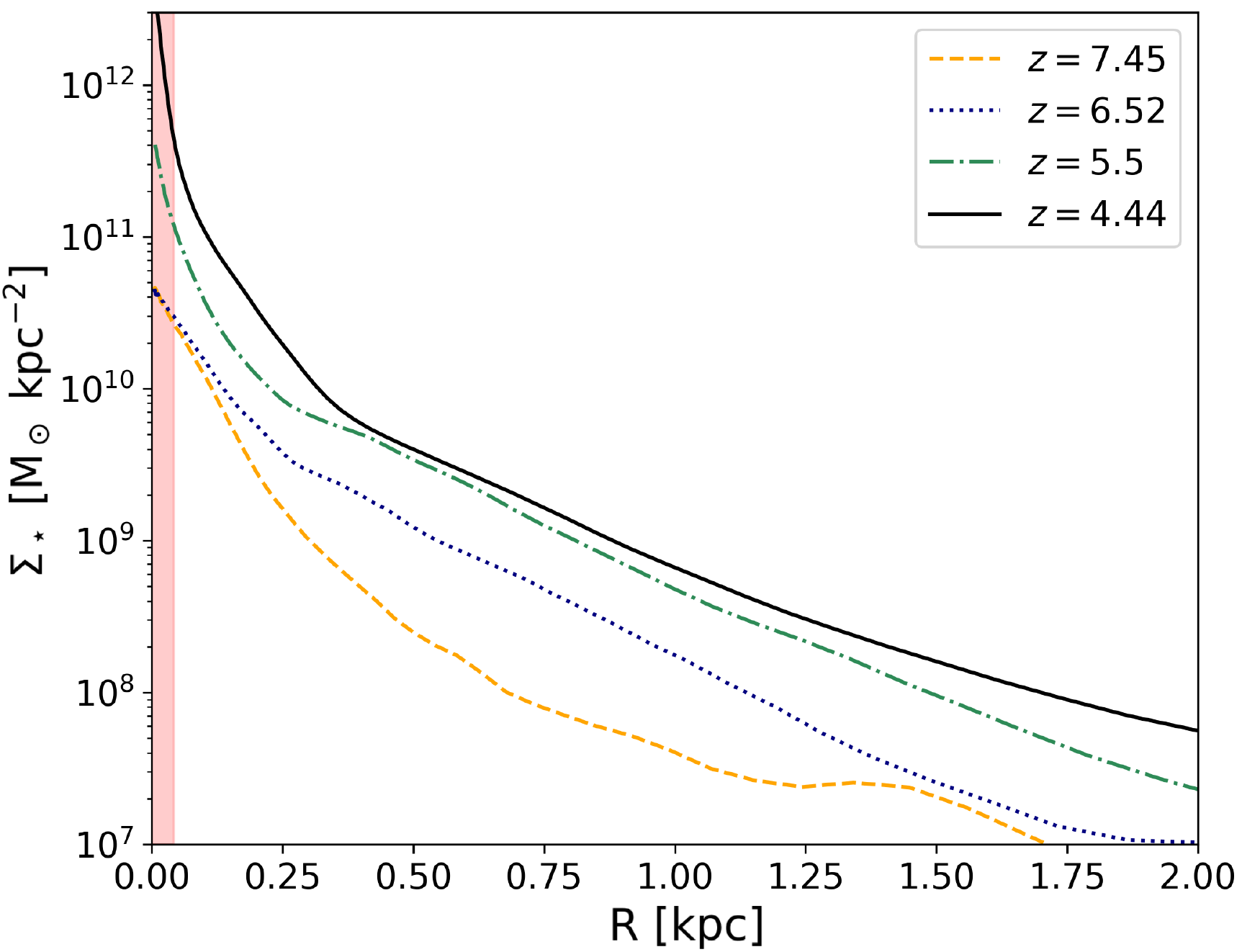}
\caption{Stellar surface density profiles in the same radial range as in \citet{Fiacconi_et_al_2017}, for four different redshifts: $z=7.45$, $6.52$, $5.5$, and $4.44$. The red, shaded region depicts the softening of our simulation}
\label{fig:profiles}
\end{figure}

\begin{figure}[htp]
\centering
\includegraphics[width=0.45\textwidth]{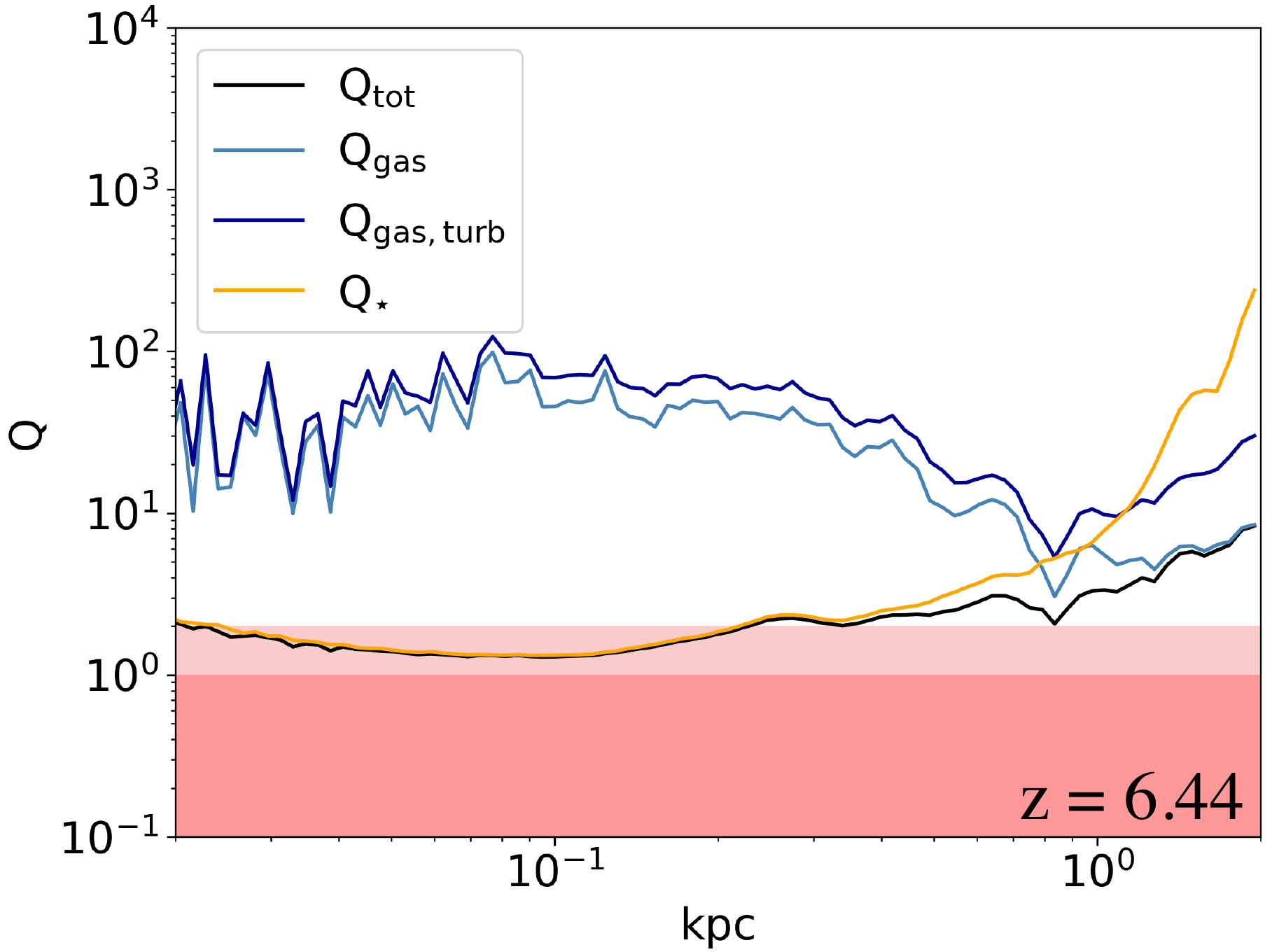}\\
\includegraphics[width=0.45\textwidth]{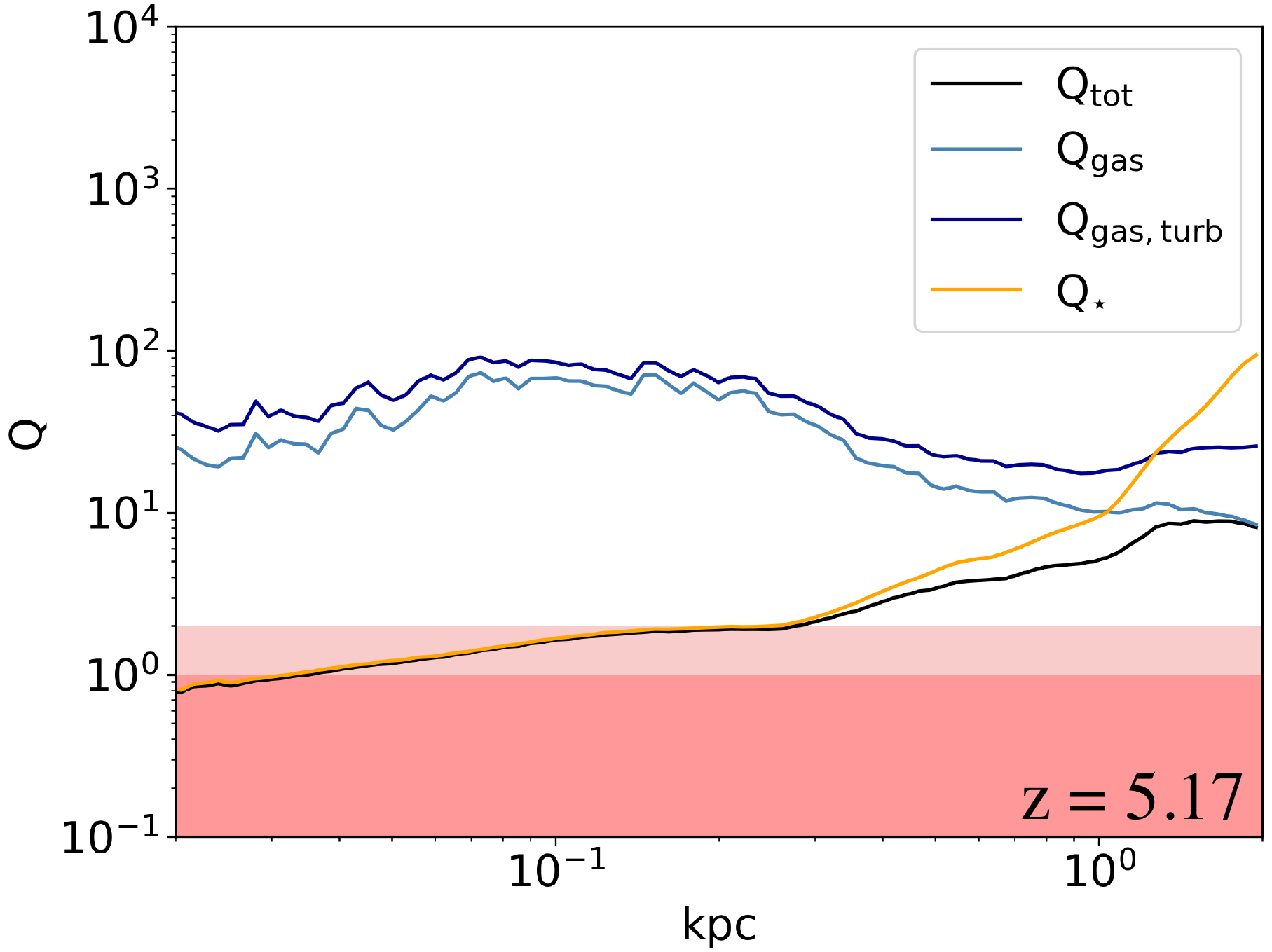}
\caption{The Toomre Q parameter of stars (\textit{orange}), gas (\textit{light blue}), and turbulent gas (\textit{dark blue}), and the total $Q$ (\textit{black}) as a function of radius at $z=6.44$ (\textit{top} panel) and at $z=5.17$ (\textit{bottom} panel). The shaded regions show the \textit{marginally stable} (pink) and the \textit{unstable} region (red). Even at $z=6.44$, which is comparable with the \textsc{PONOS-HR} data in \cite{Fiacconi_et_al_2017}, we obtain a stable stellar as well as gaseous disk.}
\label{fig:Q}
\end{figure}

The primary system is a star-forming main sequence galaxy with a relatively quiescent merger history residing in a large filament, with other converging secondary filaments (see Figure~\ref{fig:total}). In this study, we define a major merger as one with a mass ratio $q = M_{\rm sat} / M_{\rm halo} > 0.1$. In Figures~\ref{fig:mass} and \ref{fig:SFH} (\textit{top} panels; \textit{dashed} vertical lines), we show the times of the last 10 mergers, although from $z=68$ to $4.44$ we can count 14 major mergers, of which four have $q>0.25$. At the final redshift, $z=4.44$, the main halo has a virial radius of $37$~kpc, a virial DM mass of $2.3 \times 10^{11}$~M$_\odot$, a virial stellar mass of $3.1 \times 10^{10}$~M$_\odot$, and a virial gas mass of $1.3 \times 10^{10}$~M$_\odot$ (from the AHF output file). In Figure~\ref{fig:mass}, we also show the distribution of the enclosed DM mass, which shows a steep rise in the central region, although the total potential in the center is dominated by the stellar contribution out to $\approx 30$~kpc. The galaxy grows quickly in mass, with its total mass being already well above $10^{11}{M_{\odot}}$ at $z \sim 6.5$. Interestingly, at this point the mass is comparable to that of the main galaxy in another zoom-in run performed with nearly identical hydrodynamical solver and sub-grid physics, the \textsc{PONOS-HR} simulation described in \citet{Fiacconi_et_al_2017}. However, the halo in that simulation, by the present time, is nearly six times more massive than that of \textsc{GigaEris} (an early type galaxy, rather than a spiral, forms in a lower-resolution version of the \textsc{PONOS-HR} simulation by low redshift, see \citealt{Ardila_et_al_2021}). The baryonic matter increases significantly its relative weight in the mass budget within the inner 10--20~kpc between $z \sim 7$ and $z \sim 4$. In Figure~\ref{fig:SFH}, we show that our galaxy lies on the main sequence (see \textit{top} panel) and that the star formation history is in good agreement with observations from the VIMOS Ultra-Deep Survey \citep[VUDS;][]{Tasca_et_al_2015}, which is a spectroscopic redshift survey of faint galaxies mainly in the redshift range $4.5<z<5.5$ (see \textit{bottom} panel). Furthermore, we also show the result from \textsc{PONOS-HR}, as a blue square (see \textit{bottom} panel), which highlights the similarities between these two simulations.

\begin{figure*}[htp]
\centering
\includegraphics[width=1\textwidth]{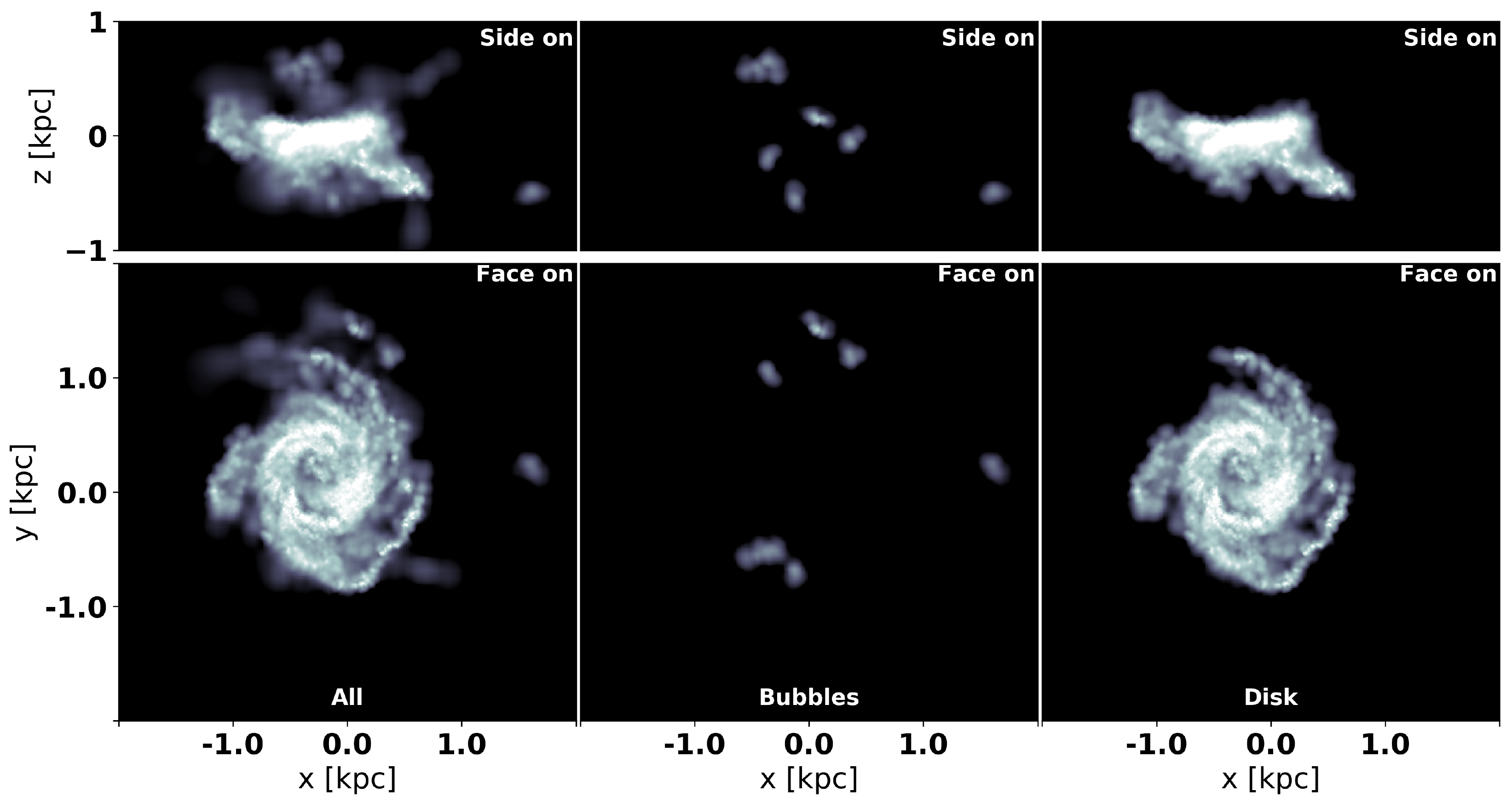}
\caption{Stellar surface density ``face on'' (\textit{bottom} panels) and ``edge on'' (\textit{top} panels), with the total stellar angular momentum vector pointing along the $z$-axis, of the main \textsc{GigaEris} galaxy at $t=0.48$~Gyr after the Big Bang ($z=6.95$). The figure is divided in three vertical panels. \textit{From left to right:} all newly born stars in that snapshot; only the stars born in individual groups (``\textit{Bubbles}''); and the disk stars.}
\label{fig:split}
\end{figure*}

\begin{figure*}[htp]
\centering
\includegraphics[width=0.85\textwidth]{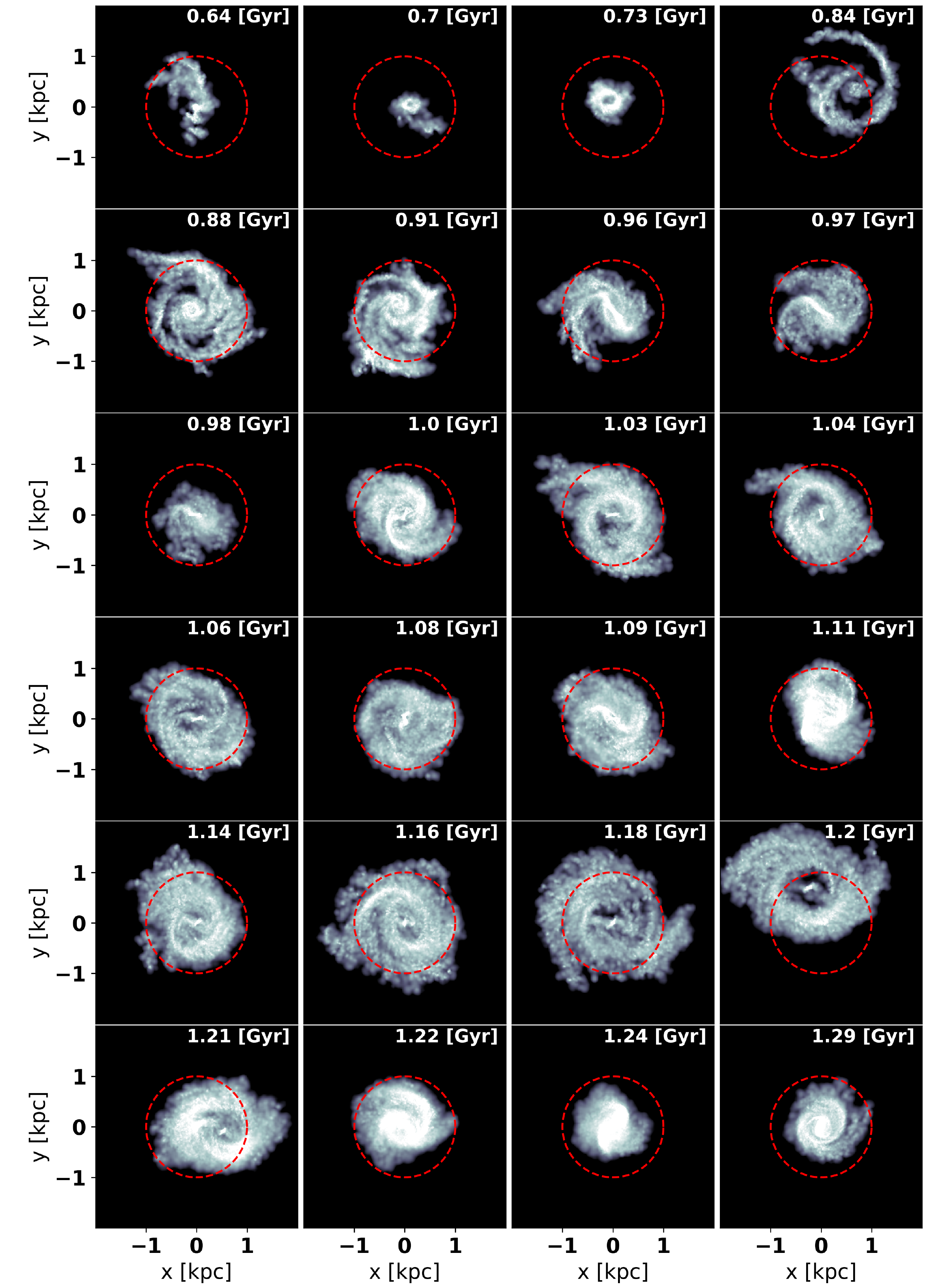}
\caption{Surface density of the newly born stars (face on) of the main \textsc{GigaEris} galaxy as a time sequence. To make the evolution of the disk clearly visible, we added a red circle with a radius $r =1$~kpc.}
\label{fig:split_time_evo_faceon}
\end{figure*}

\begin{figure*}[htp]
\includegraphics[width=1\textwidth]{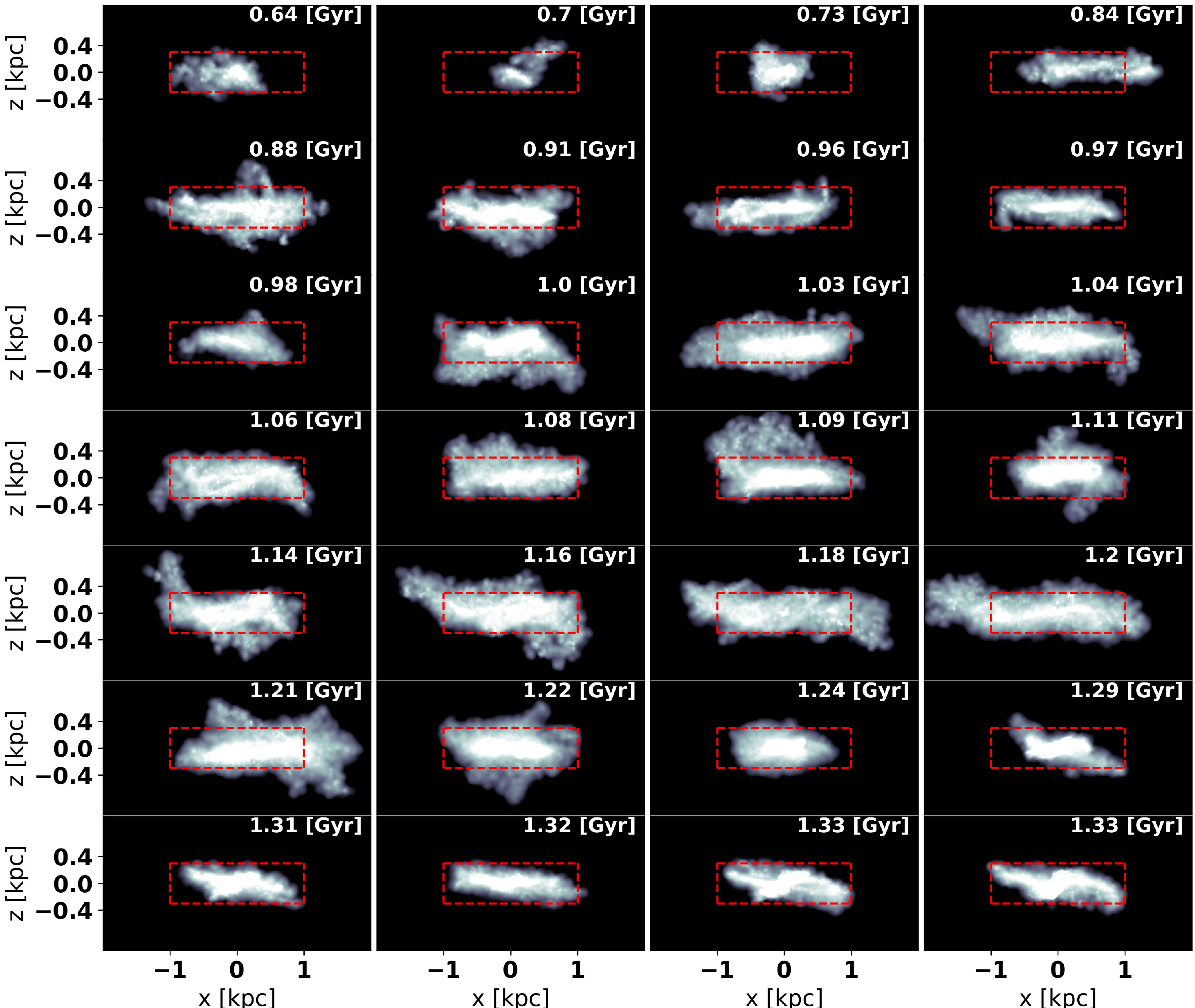}
\caption{Surface density of the newly born stars (edge on) of the main \textsc{GigaEris} galaxy as a time sequence. The thin disk appears to exist in every snapshot, although we can observe a slight warp in some cases. The red box indicates a height of $600$~pc and a width of $2$~kpc.}
\label{fig:split_time_evo_edgeon}
\end{figure*}

\begin{figure*}[htp]
\centering
\includegraphics[width=0.8\textwidth]{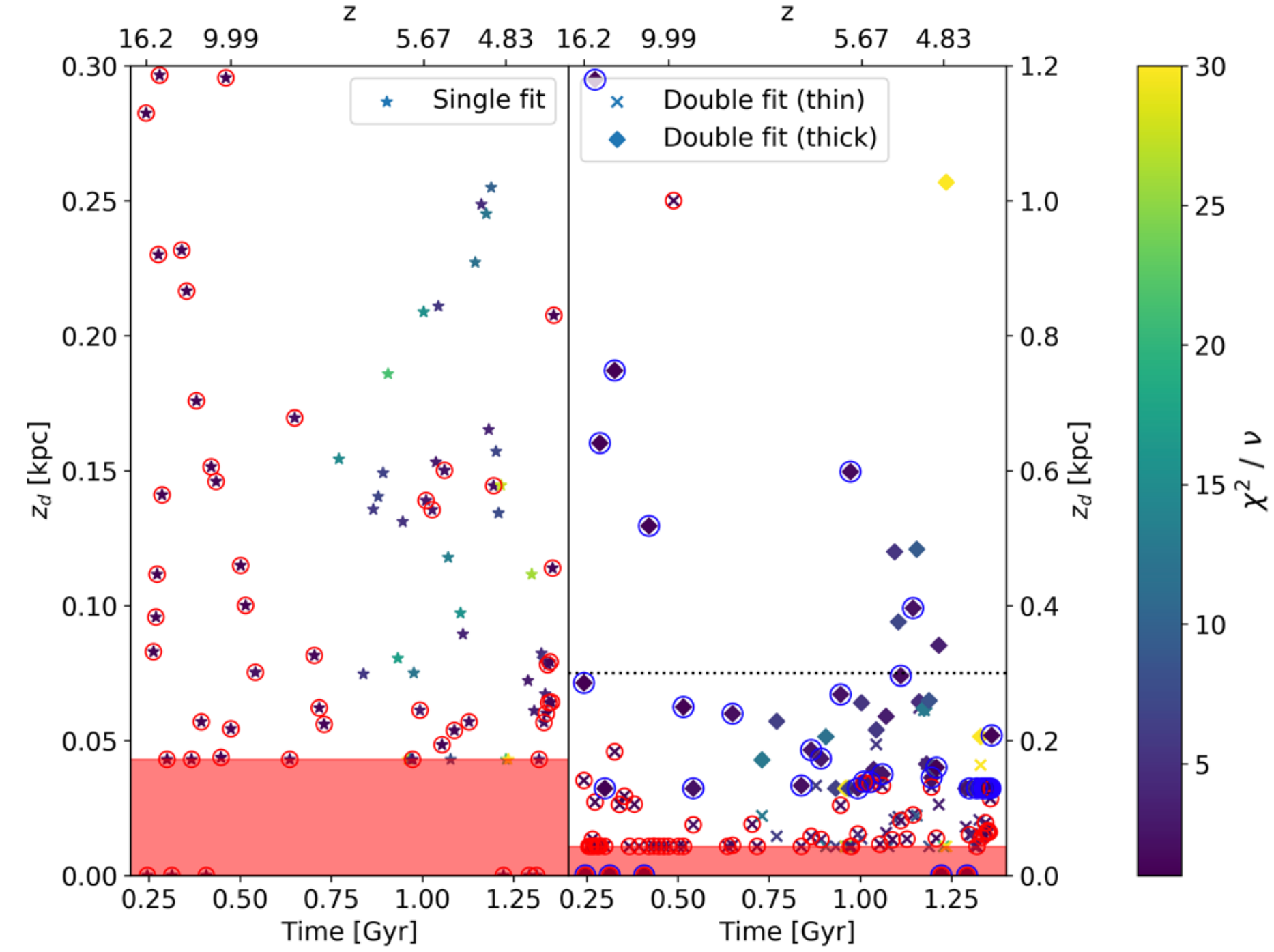}
\caption{Characterization of disk thickness for newborn disk stars in various snapshots as a function of time. The \textit{left-hand panel} shows the results of the single profile fitting procedure to infer the height $z_{\rm d}$ using 80$\%$ of the enclosed mass. The \textit{right-hand panel} depicts the outcome of the double profile fitting procedure to infer the scale height $z_{\rm d}$ (\textit{crosses} and \textit{rhombuses} for the thin and thick component, respectively), again at $80\%$ of the enclosed disk mass. The encircled values have a reduced $\chi^2$ value between 1 and 3. The red shaded area indicates the softening length of the simulation and the black, horizontal dotted line on the right-hand side shows the maximum range of the left-hand plot.}
\label{fig:fit_results}
\end{figure*}

\subsection{Disk formation and dynamics of newly born stars}

\begin{figure*}[htp]
\centering
\includegraphics[width=0.95\textwidth]{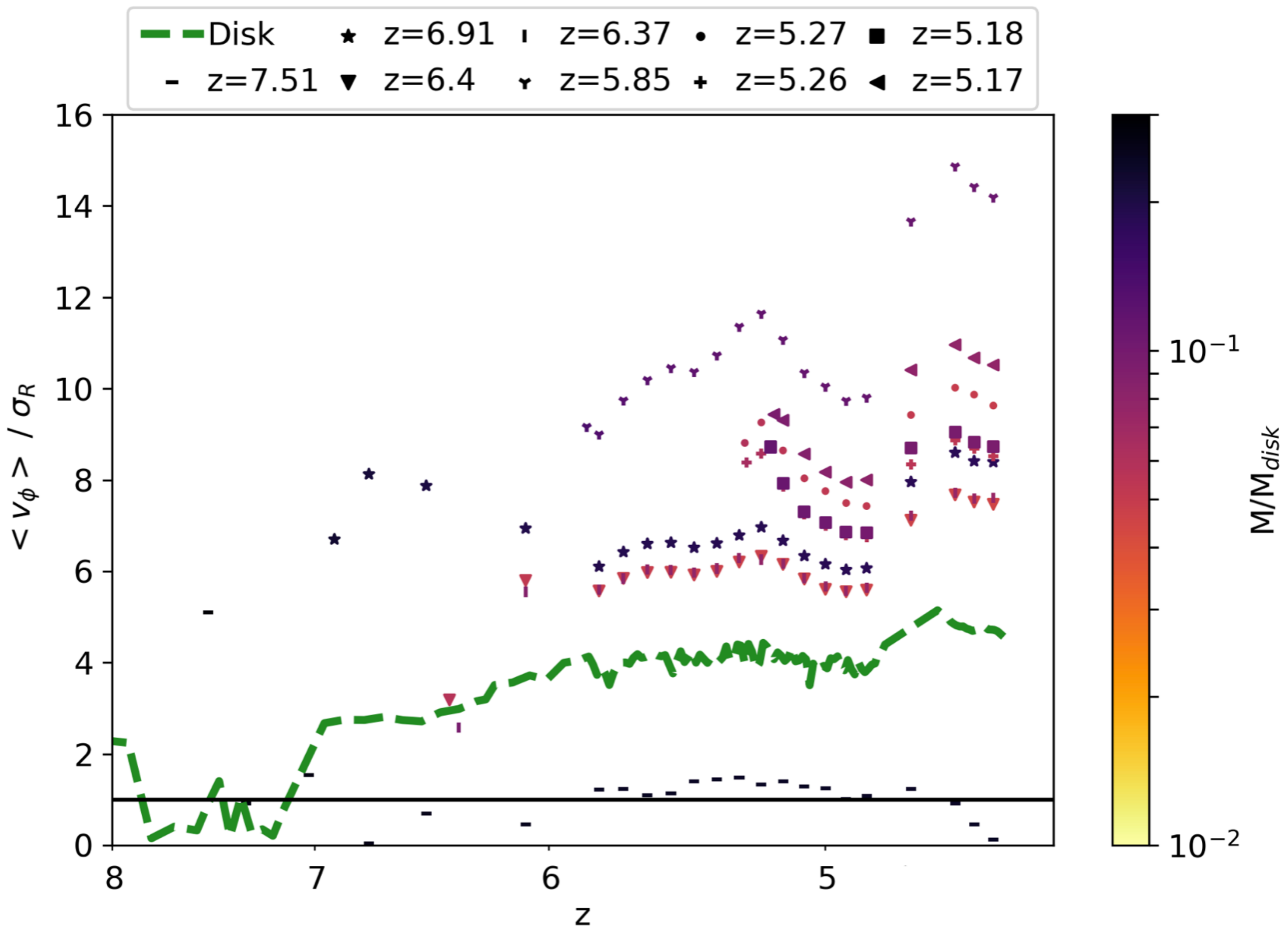}
\caption{The figure shows the kinematical evolution of coeval stellar disk sub-components identified by the \textsc{DBSCAN} algorithm. For any component, labelled by a birth time in the figure, the same stars are followed until the final redshift, hence they are coeval. We show their $<v_\phi>/ \sigma_{R}$ values as a function of time. The stars are tracked throughout the simulation and the associated disk is always realigned such that the angular momentum vector points towards the $z$-direction. The scatter-points are color-coded by the ratio of the cohorts mass to the total disk mass at the corresponding time step. It is obvious that the $<v_\phi>/ \sigma_{R}$ does not decrease as in Figure \ref{fig:vcirc}, but rather increases. With this method, we are not biased by ex-situ formation and galaxy mergers and therefore obtain a larger $<v_\phi>/ \sigma_{R}$ value. The green line depicts the value for the total galactic disk and it can be seen that this value is always significantly lower than that of the individual coevolving disks. We also obtain high (and stable) values of $<v_\phi>/ \sigma_{R}$ as early as at redshift $z=7$.}
\label{fig:v_sigma_tracking}
\end{figure*}

A first remarkable finding can be already seen in Figure~\ref{fig:ALL_Bird2}, which shows the stellar surface density maps, which are divided into four rows and four columns. The maps are oriented face-on (second and fourth row) and edge-on (first and third row) on the whole stellar component of the simulation. The stellar surface densities are divided in formation time bins, analogously to the procedure described in \cite{Bird_et_al_2013}, but with a finer bin width. The interesting result is that, already by visual inspection, a flattened disk component, in which the bulk of the mass has a height below $300$~pc, is already present at $z \sim 7$--8 (see Figure~\ref{fig:ALL_Bird2}). The vertical extent of this disk is thus comparable to the present-day thin disk component of spiral galaxies, and is quantified later in Figure~\ref{fig:fit_results}. The stellar disk is rather compact in radial size at this redshift, though, as it extends out to no more than 1~kpc. The compact size reflects the small virial radius of the host halo, which indirectly imposes a ceiling on the orbital angular momentum of baryonic matter that can be gravitationally bound to the halo and cool down to form a disk \citep[][]{White_et_al_1978, Efstathiou_et_al_1980}. The overall aspect ratio of the disk (scale height/disk scale length) is therefore larger than in present-day counterparts. The newborn stars have a similar aspect ratio even at this early time, but this becomes increasingly smaller, with the ratio between vertical and radial scale length reaching a value $< 10$ by $z < 5$, more akin to the thin disk in present-day spirals (Figures~\ref{fig:ALL_Bird2} and \ref{fig:ALL_Bird}). This thin disk, due to our limited time evolution, may or may not be the seed of the galactic scale disk that the \textsc{Eris} system eventually develops. It is possible that such a thin disk might form and subsequently be disrupted during the early course of galaxy evolution. But since our galaxy has a rather quiescent merger history, with no major merger after $z=3$, we come to the conclusion that this is indeed the origin of today's Milky Way's thin disk. Furthermore, a more extended stellar envelope surrounding the thin disk grows with time (see the transition from Figure~\ref{fig:ALL_Bird2} to Figure~\ref{fig:ALL_Bird}, which shows the surface densities at $z=4.44$). In the following subsections, we will characterize the galaxy by means of its global properties, including their time evolution, trying to shed light on the origin and evolution of this ubiquitous thin-disk component.

We can now characterize the state of the stellar disk more in-depth, and compare it with previous work, such as \cite{Bird_et_al_2013, Bird_et_al_2021}, by comparing the lower panels in Figure~\ref{fig:ALL_Bird}, in which we observe the same \textit{``inside-out''} and \textit{``bottom-up''} growth as previously shown by \cite{Bird_et_al_2013}. However, conversely to previous work, we find that there is always a high central density with a thin component that can be attributed to a young stellar disk. Furthermore, during the entire simulation, the density maps always exhibit a thin disk, and  portray prominent morphological features such as a bar or a rotating disk. The latter seems less prominent at the final snapshot, though these features can be easily seen at other times, such as at $z=6.95$ (Figure~\ref{fig:ALL_Bird2}). 

To further demonstrate the existence of a thin disk even at high redshifts, we study the kinematics of the stars. We compute mean values of the rotational velocity $v_{\phi}$ and of the radial velocity dispersion $\sigma_{R}$. In particular, in Figure~\ref{fig:vcirc}, we calculate the ratio of rotation to velocity dispersion $<v_\phi>/\sigma_{R}$, which is a measure of how prominent a kinematically cold rotating disk of stars is. For this purpose, we consider particles in a cylinder with $r<2$~kpc and $-1$~kpc $ < z < 1$~kpc, binning them using the same formation time bins as in the surface density maps. We observe that, for all age cohorts, the profile peaks at a high value above unity and then decreases as time progresses. In addition, profiles for stars selected with increasingly younger ages peak at progressively larger values, reaching even above 5, as expected from \textit{``inside-out''} disk formation bringing in progressively higher angular momentum baryons as more time elapses  \citep[][]{Bird_et_al_2013, Sokolowska_et_al_2017}. This is consistent with the low-redshift analysis of stellar kinematics for stars of different age cohorts made by \cite{Bird_et_al_2013,Bird_et_al_2021}. 

We find that, concurrently, the rotational velocity itself always increases with time, growing from less than $100$~km~s$^{-1}$ at $z \sim 10$ to more than $300$~km~s$^{-1}$ towards $z \sim 4$. Therefore, the fact that the $<v_\phi>/\sigma_{R}$ tends to decrease with time for the individual age cohorts, as shown in Figure~\ref{fig:vcirc}, must be due to an increase in stellar velocity dispersion. The latter can be attributed to a variety of agents. Both accretion of stars with similar age but hotter kinematics, occurring through the many minor mergers impinging on the galaxy after $z=10$ (see Figure~\ref{fig:mass}), and heating of in-situ stars by internal dynamical instabilities and perturbations by incoming massive satellites, can be responsible. 

Among the internal heating mechanisms, there are disk thickening due to bar formation, which could play a role as a strong bar is seen to develop at $z < 6$ (the formation and evolution of a strong bar will be the topic of an upcoming paper), and various formation planes. The development of the bar is reflected in the evolution of the stellar surface density profile shown in Figure~\ref{fig:profiles}. Indeed, at $z < 6$ the profile becomes much steeper inside 0.5~kpc, which corresponds roughly to the extent of the bar. This reflects outward transport of angular momentum by the bar, which generates a nuclear gas inflow and nuclear star formation \citep[see e.g.][]{Debattista_Mayer_et_al_2006, Guedes_et_al_2013}. This result is at variance with previous high-resolution simulations at high redshift, such as \textsc{PONOS-HR} \citep[run down to $z = 6.5$;][]{Fiacconi_et_al_2017}, which exhibited nearly perfect exponential profiles. We caution, however, that bar formation is very sensitive to slight changes in the potential, as shown by the appearance of the bar in another version of the \textsc{PONOS-HR} simulation run to lower redshift \citep[][]{Bortolas2020}. In order to quantify bar-driven heating or other internal heating mechanisms, though, the first step is to quantify, in general, how significant is the heating of in-situ stars, namely separating out the presumably kinetically hotter stars that are gradually added as accretion and merging continues. This will be studied in detail in the next sub-section, in which we will present a robust way to identify in-situ disk components.

As we have just discussed, from the $<v_\phi>/\sigma_{R}$ values (see Figure~\ref{fig:vcirc}) we have clear evidence of a kinematically cold disk at any redshift. In order to further characterize the physical properties of this, we now compute the Toomre Q parameter \citep{Toomre_1964}. We do this for both the stellar and gas particles within a cylinder of 2~kpc height and radius of 2~kpc:

\begin{equation}
    Q = \frac{\kappa V}{A G \Sigma},
\end{equation}

\noindent with $\kappa = \sqrt{2(v_\phi/R)^2 (1+d \log v_\phi/d \log R)}$, $A = A_{\rm g} = \pi$ and $A = A_\star = 3.36$, $V = V_\star = \sigma_{R}$ and $V = V_{\rm g} = c_{\rm S}$ (or, for turbulent gas, $V = V_{\rm g} = \sqrt{c_{\rm S}^2 + \sigma_{{\rm g},R}^2}$ (where $c_{\rm S}$ is the speed of sound). Furthermore, we also corrected for disk thickness, by multiplying Q by

\begin{equation}
T = \begin{cases} 
1 + 0.6(\sigma_z/\sigma_R)^2  & {\sigma_z/\sigma_R} < 1/2,\\
0.8 + 0.7(\sigma_z/\sigma_R) & \sigma_z/\sigma_R \geq 1/2,
\end{cases}
\end{equation}

\noindent where we followed \citeauthor{Romeo_et_al_2011} (\citeyear{Romeo_et_al_2011}; see also \citealt{Romeo_1994}; \citealt{Romeo_et_al_2013}; \citealt{Inoue_et_al_2016}). We can write the Toomre Q parameter as

\begin{equation}\label{eq:Q}
Q_{\rm tot}^{-1} = \begin{cases} 
WQ_\star^{-1} +Q_{\rm g}^{-1}  & Q_\star \geq Q_{\rm g},\\
Q_\star^{-1} + W Q_{\rm g}^{-1}  & Q_\star <  Q_{\rm g},
\end{cases}
\end{equation}

\noindent with $W = 2V_\star V_{\rm g} /(V_\star ^2 + V_{\rm g} ^2 )$ \citep[again, see][]{Romeo_et_al_2011}. In Figure~\ref{fig:Q}, we show the Toomre Q parameter and conclude that we have a stable stellar disk at $z\approx 6.44$, which is roughly the same time of the analysis in \textsc{PONOS-HR} \citep[][]{Fiacconi_et_al_2017}. The Toomre Q, however, is higher in \textsc{PONOS-HR} (see Discussion below). Furthermore, at $z < 6$ the Q of the stars decreases significantly as the disk becomes unstable to bar formation, which in \textsc{PONOS-HR} \citep[run down to $z = 6.5$;][]{Fiacconi_et_al_2017} is not observed. This likely reflects the higher stellar mass and relatively lower halo mass of \textsc{GigaEris}, which renders its disk more unstable to non-axisymmetric instabilities.

\subsection{A Multi-component primeval disk galaxy}

Analyzing how and where star formation proceeds as the galaxy evolves led us  to the conclusion that stars do not form only in a single plane or layer, but rather in multiple sub-units having various heights, radial extents, and relative orientations. To properly define a (thin) disk  and  other  accompanying  components,  we  applied the \textsc{DBSCAN} \citep[][]{DBSCAN} clustering algorithm to our data set. The algorithm, which can be described as a special case of the Friends-of-Friends algorithm, as shown by \cite{Kwon_et_al_2010}, identifies the main disk body by using a density threshold  to  group particles (we choose a density threshold of roughly $1$ M$_\odot$pc$^{-2}$ within 100 pc around each star), and then discards particles that are spatially separated from any of the identified groups. More specifically, a star  particle  belongs  to  a  group  if it encompasses 40 other stellar particles within a sphere of radius $100$~pc. Particles that have fewer star particles than the nominal 40 within their own sphere, but are inside the sphere of another particle that satisfies the criterion, are also accounted for. This way we identify sequences of mutually spatially connected particles, with a procedure conceptually similar to that of a gather/scatter kernel interpolation in SPH. Particles that do not define a group, or do not overlap with any group, are defined as isolated, and discarded as background noise. We define the largest coherent group as the disk. The algorithm also finds other groups, which are not mutually connected with such largest group, which typically correspond to diffuse, three-dimensional features above and below the disk defined by the largest group. These are ``bubbles'' of extraplanar star formation (see this procedure in Figures~\ref{fig:split}). Note that such bubbles are well resolved, encompassing up to about 4000 particles each. We remark that these pockets of extraplanar star formation would be barely resolved in conventional zoom-in simulations, which have a mass resolution more than one order of magnitude lower than that of \textsc{GigaEris}, and would then be discarded as background noise. We then proceed to quantify the structure of the disk of newborn stars discarding both the noise (individual particles) and the extraplanar bubbles, which will be studied in a future paper dedicated to diffuse baryonic components in the simulation.

In summary, we can outline the overall  procedure to identify the newborn disk as follows:

\begin{itemize}

    \item Select a large box around the center of the galaxy [$-4$~kpc $< x < 4 $~kpc, $-4$~kpc $< y < 4 $~kpc, $-4$~kpc $< z < 4 $~kpc].
    
    \item Find the newborn stars within this region [d$t = 0.017$~Gyr].
    
    \item Apply the \textsc{DBSCAN} algorithm to identify the main disk body as described above, discarding background noise and bubbles.
    
    \item Recenter on the remaining disk stars and orient the disk such that the angular momentum vector is perpendicular to the $xy$-plane (``face on'').
    
\end{itemize}

This procedure can be applied recurrently to analyze the distribution and kinematics of the stars identified in each snapshot. 
We  determine the thickness via the vertical density profile at 80$\%$ of the enclosed mass estimated by

\begin{equation}
\rho (z) =  C  {\rm sech}^2 \left( \frac{z-\mu} {z_{\star} } \right),
\end{equation}

\noindent with $C$ being a normalization constant, $\mu$ the mean value of the vertical component of the disk, and $z_\star$ the disk thickness. Additionally, since we already know that the disk is not formed in a single disk-like layer but instead within a complex structure, we also performed a two-component ${\rm sech}^2$ profile fit:

\begin{equation}\label{eq:doublesech}
\rho (z) =  C_1  {\rm sech}^2 \left( \frac{z-\mu} {z_{\rm thin} } \right) + C_2  {\rm sech}^2 \left( \frac{z-\mu} {z_{\rm thick} } \right),
\end{equation}

\noindent with $C_1$ and $C_2$ being normalization constants, $\mu$ again the mean $z$-value of the disk stars, and $z_{\rm thick}$ and $z_{\rm thin}$ the thick and thin disk scale heights, respectively. In Figure~\ref{fig:fit_results}, we show the results of both fitting procedures. With this method, we are always able to find a thin disk component independent from a single or double fit. The left-hand side clearly depicts that a single fit yields very good $\chi^2/\nu$ values\footnote{The $\chi^2$ errors have been estimated by assuming a Poisson sampling noise.} (smaller than 3) and that all are below $300$~pc. Therefore, we can conclude that a thin disk is always in existence, but can be hidden by ex-situ formation or stellar accretion via consecutive galaxy mergers. We also want to mention that a two-component fit (Equation~\ref{eq:doublesech}) will likely lead to degenerate results, but since our results from both fits are consistent with each other and always display a thin disk, we do not further investigate this issue.

We then revisit the evolution of the kinematics of in-situ  stars by tracking them over time. Specifically, we identify ensembles of stars that are born at a given time in the disk, identified by \textsc{DBSCAN}, measure their $<v_{\phi}>/\sigma_R$ at the initial time, and track these same stars till the end of the simulation. By repeating the procedure at each snapshot, we identify a set of coeval sub-components of the disk (coeval disks). The dynamics of the stars in these sub-disks can be perturbed by satellites or internal instabilities such as bars, but there is no contribution of ex-situ stars by construction, given the way we identify the coeval sub-disks. Therefore, in Figure~\ref{fig:v_sigma_tracking}, we show the time evolution of  $<v_\phi>/\sigma_R$ for multiple co-evolving disks identified with \textsc{DBSCAN}. In contrast to Figure~\ref{fig:vcirc}, we can see that once an individual sub-disk is formed it remains a kinematically cold structure, hence it is not heated or destroyed by any of the aforementioned mechanisms. On the contrary, quite surprisingly the ratio $<v_\phi>/\sigma_R$ increases with time in all cases. Moreover, the $<v_\phi>/\sigma_R$ of the newly born stars (Figure~\ref{fig:v_sigma_tracking}) is always higher than the $<v_\phi>/\sigma_R$ of the global galactic disk (Figure~\ref{fig:vcirc}), which includes the contributions from all the coeval disks. The increase is largest at the latest times. This suggests that what we are witnessing is indeed higher angular momentum disk components building up with time, a well established fact in cold DM (CDM) cosmogonies \citep[e.g.][]{Sokolowska_et_al_2017}, but also that, as such new higher angular momentum components assemble, the stars belonging to previous coeval disks are not heated up significantly, rather they increase their rotation speed as more mass is added in the disk itself, because kinetic energy has to increase if the system has to remain in equilibrium.

\noindent Overall the picture that emerges is that the decrease of $<v_\phi>/\sigma_R$ shown in Figure~\ref{fig:vcirc} is nearly entirely due to an increasing contribution of higher velocity dispersion stars added by accretion events, i.e. from ex-situ stars, rather than being due to heating of the pre-existing in-situ component. This essentially agrees with the notion, so far established at low redshift only, that disk heating is actually a rather mild dynamical driver of disk evolution, contrary to claims in early studies of galaxy-satellites interactions \citep[][]{Kazantzidis_et_al_2009, Grand_et_al_2016}. The fact that, in Figure~\ref{fig:v_sigma_tracking}, the overall disk  has a milder increase of $<v_\phi>/\sigma_R$ reflects the fact that the angular momenta of the sub-disks are not aligned, so that, when they are superimposed, the net angular momentum around the mean rotation axis can only be smaller than that of the individual components (note that also for the global disk  we are only considering in-situ stars).

\section{Discussion}

\begin{figure*}[htp]
\centering
\includegraphics[width=\textwidth]{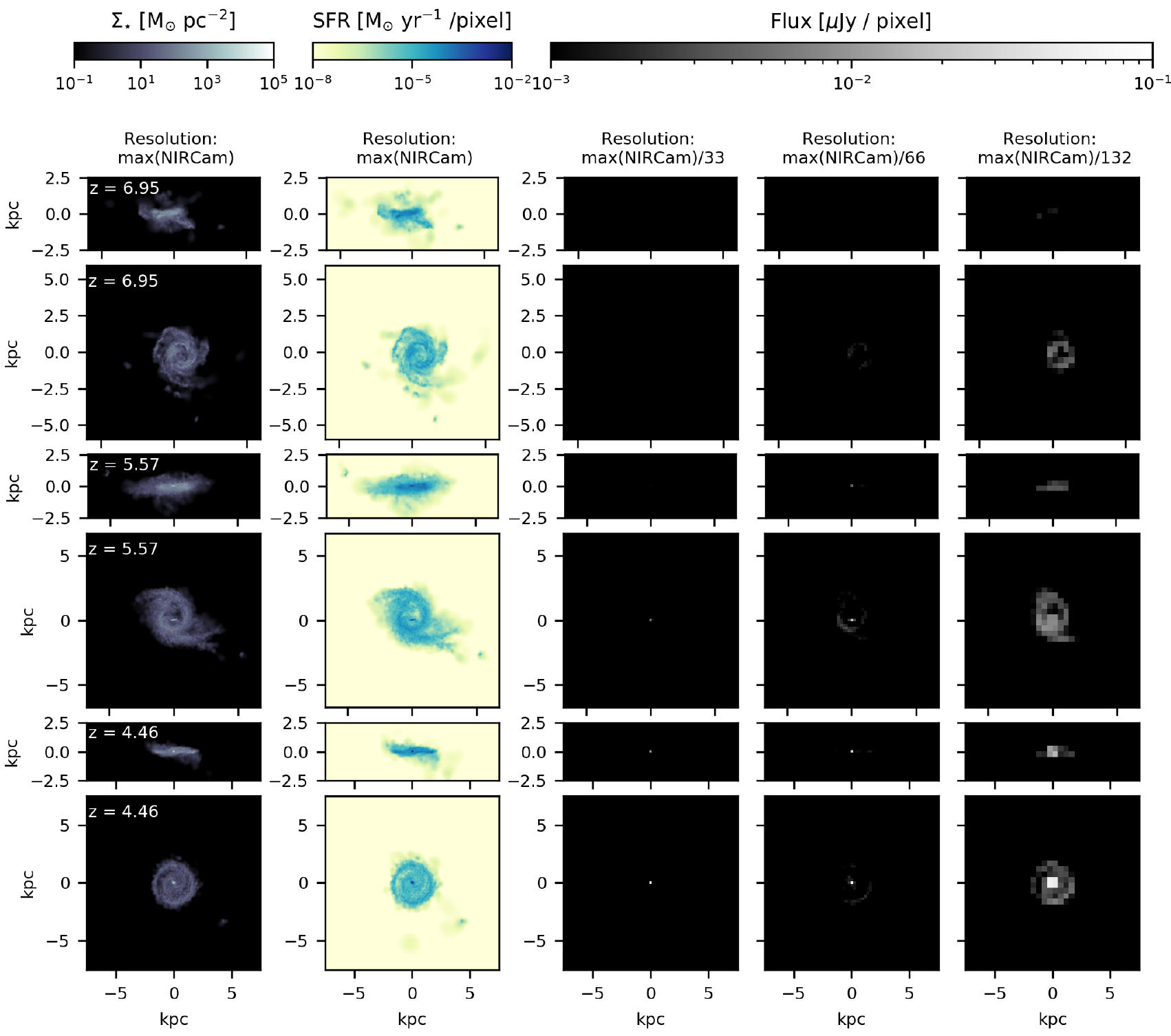}
\caption{Mock observations with the NIRCam camera of JWST at short wavelength imaging. On the left-hand side we show the theoretical surface density of the galaxy at three different redshifts, $z=6.95, 5.57$, and $4.46$ (\textit{from top to bottom}), with a spatial resolution comparable to the NIRCam resolution. The \textit{second} column shows the star formation rate per resolution element (``pixel''). Afterwards, we transform the star formation rate to $\mu$Jy using the relations given by \cite{Kennicutt_1998} and combine as many pixels as necessary to obtain a reasonable flux (see \textit{last three} columns). In our example, we used an exposure time of 7 cycles (roughly 3h 49min) and a signal-to-noise ratio of roughly 5 at $10^{-2}$~$\mu\text{Jy}$. We can see that all disks, which are depicting newborn stars, can be reasonably resolved at all redshifts. Therefore, JWST will be an ideal instrument to find high-redshift disk galaxies.}
\label{fig:JWST2}
\end{figure*}

\begin{table}[]
\centering
\begin{tabular}{ccccc}
 Name & Wavelength & Area & Resolution & Geometry \\
 & [$\mu$m] & [arcmin$^2$] & [$"$/pix] & [$"$x$"$]\\
 \hline
MIRI  &  5.6-26.5 & 3.1 & 0.11 & 74 x 113  \\
NIRCam:& &  & & \\
lw & 2.4-5.0 & 9.7 & 0.063 & 264 x 264\\
sw & 0.6-2.3 & 9.7 &  0.031 & 264 x 264 \\
NIRISS  & 0.8-5.0 & 4.84 & 0.065 & 133 x 133  
\end{tabular}\label{tab:JWST}

\caption{The four cameras of JWST (Miri, NIRCam-lw, NIRCam-sw and NIRISS) and their wavelength bands, square area, resolution per pixel, and geometry.}
\end{table}

In this work, we have shown that stars are assembled in a flattened disk component already at very high redshift, as early as $z\sim 8$, in a progenitor of a present-day Milky Way-sized galaxy halo. A thin disk is present till the end of the simulation, at $z \sim 4.4$, and results from subsequent episodes of star formation from newly accreted cold gas. 

The implication is that, unless dynamical heating, internal or external, heats up considerably the entire early thin disk over the following several Gyr, a remnant old thin disk component, with age $> 10$~Gyr, should be present in present-day massive spirals. Indeed this is in good agreement with new observations from the Gaia DR2 release, in which roughly 20 to 30$\%$ of the oldest stars of the Milky Way, with a [Fe/H] $\leq -2.5$ and thus born only 2--4~Gyr after the Big Bang \citep[see][]{El_Badry_et_al_2018}, are arranged in a thin rotating disk \citep[][]{Sestito_et_al_2020}. The comparison with the present-day thick disk of the Milky Way is instead not possible, as both internal dynamical processes and accretion will occur from $z \sim 4$ to $z=0$, which we cannot capture due to the limited evolutionary timescales probed by \textsc{GigaEris}. Therefore, it cannot be excluded that at least part of the primeval thin disk will contribute to the thick disk at a later time, or that it might evolve into a pseudoubulge via disk instabilities \citep[][]{Guedes_et_al_2013}, which would match its compact radial size scale.
 
We also find that only a fraction of the stars that are present at redshift $z=4.44$ are actually born in-situ, the vast majority of the stars being born outside of the final ``today's'' stellar disk. We speculate that the primeval thin disk has been missed in former studies due to the latter fact combined with limited mass and spatial resolution. We argue that the \textsc{GigaEris} simulation portraits evidence that the formation mechanism of high-redshift disks is fundamentally analogous to that of low-redshift disks. With enough resolution to resolve the flow in the circumgalactic and interstellar medium down to scales of a few pc, radiative cooling does produce thin rotationally supported layers, out of which stars form, already at very high redshift, and naturally build up a stellar component in a relatively thin disk. A major difference with low redshift is that, due to the much smaller halo scale size, which reflects the nature of hierarchical structure formation in CDM, the angular momentum of accreting baryonic matter is lower at high redshift, resulting in an initially smaller disk aspect ratio. Yet the stellar kinematics that result are typical of a thin disk, rather than of a thick disk or spheroid, already then. Another difference relative to low-redshift disks is that, owing to shorter infall times, filamentary accretion more frequently assembles sub-components of the disk whose planes have mutual inclinations. However, as shown by our \textsc{DBSCAN} analysis, these different coeval disks all have cold kinematics and, at least in the object under study, globally the kinematics is still that of a rotationally supported disk, albeit with somewhat hotter kinematics than some of the individual sub-components.

Since only a small fraction of the final stellar disk (at redshift $z=4.44$) is made up of in-situ stars, we further theorize that the bulk of the thick disk is created either due to constant accretion and (minor) merger events as described in \cite{Minchev_et_al_2015} or due to internal heating mechanisms as described by \cite{Park_et_al_2021}. Nonetheless, the build-up and origin of the thick stellar disk will be the subject of future work.

Another zoom-in simulation with comparable mass and force resolution, and similar sub-grid physics, \textsc{PONOS-HR}, which follows the formation of a nearly one-order-of-magnitude more massive system, revealed a thicker disk, with a hotter interstellar medium, driven by powerful supernovae explosions. At a comparable redshift, $z \sim 6.5$, \textsc{GigaEris} has indeed a lower Toomre Q parameter, even when accounting for gas turbulence (Figure~\ref{fig:Q}), relative to \textsc{PONOS-HR} (see figure~$9$ in \citealt{Fiacconi_et_al_2017}), and its stellar surface density profile is also significantly steeper in the inner kpc region, whereas \textsc{PONOS-HR} exhibited an almost perfectly single exponential profile throughout (see Figure~\ref{fig:profiles}). Note also that the stellar mass is a factor of two larger in \textsc{GigaEris} at this time, despite the fact that the halo has a virial mass a factor of two lower compared to \textsc{PONOS-HR}. Overall the stellar disk in \textsc{GigaEris} is thus both more massive and more centrally concentrated. This suggests, thus, significant variation in galactic structure in galaxies that, at high redshift, have a rather similar stellar mass. Such diversity is likely due to different halo assembly history and  cosmic web environment generating a different gas infall/accretion flow. Overall, it would seem that in \textsc{PONOS-HR} feedback has been more effective at suppressing the accumulation of baryons in the central region of the halo, thus maintaining a lower ratio of the stellar-to-halo mass. Nevertheless, even in \textsc{PONOS-HR} a disk component is clearly dominant, albeit more turbulent and kinematically hotter than in \textsc{GigaEris}. Whether these differences in the early disk assembly are typical between systems that are progenitors of massive spirals as opposed to massive isolated early-type galaxies cannot be assessed until a similarly detailed structural study will be possible in a larger sample of simulations. Another important aspect that will have to be investigated to understand the origin of these differences is the role of the warm/hot corona surrounding the early disk. This is born out of both shocks in the accretion flow and feedback. \citet{Sokolowska_et_al_2018} have shown that stronger feedback leads to a faster build-up of the corona, which then partially suppresses cold flows earlier, thus limiting the growth of the baryonic disk. Major galaxy mergers have also been shown to enhance the effect of feedback, implying that the merging history of the galaxy plays a role \citep[][]{Sokolowska_et_al_2018}. Future analysis will assess if feedback, perhaps due to the different merging history,  has a stronger impact in \textsc{PONOS-HR} than in \textsc{GigaEris}, as suggested by the larger fraction of warm/hot gas ($T > 5 \times 10^4$~K) in the disk of the former. Notably, \textsc{PONOS-HR} indeed undergoes more and more recent major mergers ($q > 0.25$), indeed as many as three after $z=15$ as opposed to only one in the case of \textsc{GigaEris}, and the last one is also more recent, near $z=7$ as opposed to $z=13$ in \textsc{GigaEris}.

We also mention that we can not present any statistics on how often such a disk forms in a typical galaxy due to our small sample size. Given the resolution and particles numbers in this study it is not feasible to re-simulate this galaxy nor to focus on another galaxy within the simulated box. For larger statistics, we refer the reader to the study done by \cite{Park_et_al_2021}.

Finally, we also explored the detectability of such a high-redshift disk with the recently launched James Webb Space Telescope (JWST). In Figure~\ref{fig:JWST2}, we show the surface density maps in ideal NIRCam-sw resolution as well as the star formation surface density (see \textit{first} and \textit{second} column). We can apply the relation between the star formation rate and the luminosity in the UV band given by \citep[see][]{Kennicutt_1998}

\begin{equation}\label{eq:JWST}
    {\rm SFR} \left [\frac{\text{M}_\odot} {\text{yr}} \right] = 1.4 \times 10^{-28} L_\text{UV} \left [\frac{\text{erg}}{\text{s Hz}}\right].
\end{equation}

\noindent Afterwards, we transformed $L_\text{UV}$ to Jy and calculated the flux in $\mu$Jy:

\begin{equation}\label{eq:JWST2}
 {\rm flux} [\,\mu\text{Jy}]= (1+z) \frac{L_\text{UV}}{ 4\pi D_{\rm L}^2} \cdot 10^6,
\end{equation}

\noindent with $D_{\rm L}$ being the luminosity distance in Mpc. Furthermore, with the help of the online Exposure Time Calculator for JWST,\footnote{\url{https://jwst.etc.stsci.edu/}} we calculated that, with the NIRCam-sw camera and an exposure time of roughly 3h 49min, we can obtain a signal-to-noise ratio of 5 for a flux of $10^{-2}$~$\mu$Jy. The last three panels of Figure~\ref{fig:JWST2} show the results of summing up 33, 66 and 132 neighbouring pixels in the NIRCam-sw camera in order to obtain the necessary flux. As it can be seen, the disk would be detectable, in some cases also edge-on, not only towards the end of the simulation, at $z \sim 4.4$, but also at much higher redshift, $z \sim 7$. This opens the exciting prospect of directly testing the results of our numerical simulation. Since we have modeled a galaxy that is a typical star-forming galaxy at the relevant redshift, a non-detection of a clear disk component may be in favour of an irregular or distorted galactic structure. This could point to some issue in the sub-grid physics models, as the balance between radiative cooling and feedback is crucial to determine whether a dominant kinematically cold disk component forms or not \citep[e.g.][]{Sokolowska_et_al_2017}.

\section{Summary and Conclusion}

The main results of our high-resolution cosmological zoom-in simulation of a Milky Way-sized galaxy halo can be summarized as follows:

\begin{itemize}

    \item In our simulated galaxy a thin disk is always present since the earliest assembly stage. Such disk is kinematically cold, namely the rotation clearly dominates over the stellar velocity dispersion, as in present-day disks of spiral galaxies.
    
    \item The heating of in-situ stars by internal dynamical instabilities and external perturbations is negligible. Rather, sub-disks formed at different epochs remain kinematically cold, or even exhibit an increase of the $<v_{\phi}>/\sigma_R$ ratio. The overall decrease of $<v_{\phi}>/\sigma_R$ observed as time progresses is thus resulting primarily from the contribution of ex-situ stars.
    
    \item The disk develops a multi-component structure already early-on, as the stellar distribution thickens gradually primarily due to external perturbations or direct stellar accretion from infalling satellites. Ex-situ stars incorporated by accretion provide the main contribution to the vertically extended component of the
    disk, which is consistent with them being kinematically hotter.
    
    \item There is no \textit{``upside-down''} disk formation, rather a continuous \textit{``inside-out''} growth of a thin disk component. A stellar disk formed ``upside-down'' if the early progenitor of the disk formed hot. This happens because the star-forming gas can have a high velocity dispersion and still be Toomre unstable at high redshift, due to the high gas mass fraction. Indeed, while a thin disk with vertical extent below $0.5$~kpc is present since $z > 8$, the disk aspect-ratio is larger than in a present-day disk of a spiral galaxy as its initial radial extent is very compact (about a kpc) due to the lack of high angular momentum material at high redshift. We propose that this disk can be the seed of today's thin and old Milky Way's stellar disk.
    
    \item With the advent of JWST, as we showed in Figure~\ref{fig:JWST2}, we will be able to observe and characterize high-redshift galaxies that are formed by assembling a sequence of thin, kinematically cold stellar disks, and hence test our predictions and potentially set new constraints on the physics of galaxy formation at the earliest epochs.
    
\end{itemize}

The scenario of disk formation that emerges for the early assembly phase of a massive spiral as that modeled here seems at odds with the widely adopted ``upside-down'' disk formation model \citep[e.g.][]{Bird_et_al_2013, Bird_et_al_2021}, in which disks are born thick and the thin disk builds up later from kinematically colder, higher angular momentum gas accreting at lower redshift. However, close inspection of the evolution of thickness in the \textsc{ErisLE} run reported in \cite{Bird_et_al_2013} shows that, in the inner disk at $R < 1$~kpc, the stars older than 10~Gyr are characterized by a scale height lower than $500$~pc, while the inner disk becomes thicker with time. Since it is this inner disk less than 2~kpc in size that is in place at the redshifts considered in this paper, our results are after all not in conflict. However, due to the lack of resolution, both spatial and mass, neither \citet{Bird_et_al_2013} nor other works studied the inner disk, hence their conclusions on ``upside-down'' formation are driven by the later formation phase of the galaxy. Instead, here we clearly showed that, once numerical resolution is up to the challenge, a thin disk is seen to form even at $z \sim 7$--8, essentially as soon as the galactic disk begins to assemble. This of course has important implications for upcoming observations of the stellar component of high-redshift galaxies by JWST and other instruments. 

One may wonder which component of the present-day Milky Way, or of spiral galaxies in general, corresponds to the primeval thin disk identified here. We argue that the primeval thin disk in our simulation provides a natural explanation for the oldest thin disk component revealed by the Gaia DR2 release, but it is also possible that part of the primeval inner disk would evolve into a disky pseudobulge \citep[see also][]{Guedes_et_al_2013}, which is a ubiquitous component in the central regions of massive late-type galaxies \citep[nearly all of those in the Local Volume have such a component, see][]{Kormendy_et_al_2013}. Previous work had already found that pseudobulges could form at $z > 3$ in disk galaxies, but had attributed that mostly to the evolution of an early bar \citep[][]{Guedes_et_al_2013}. On the contrary, we believe that at least a fraction of the pseudobulge stars can originate in-situ from pre-existing stars that formed in the primeval compact, kinematically cold disk configuration. The presence of an old pseudobulge component in local late type spiral galaxies is fairly well established observationally \citep[][]{Kormendy_et_al_2019}. The latter pseudobulges, though, are often vertically extended configurations, so-called peanut-like bulges, which are consistent with bar-like components dynamically heated by a buckling instability \citep[][]{Debattista_Mayer_et_al_2006}. Whether dynamical heating occurred at low redshift for the inner disk stars in our simulation cannot be probed in \textsc{GigaEris}. It is possible that present-day pseudobulges may hide a kinematically colder component with age $>10$~Gyr. This could be tested with much more accurate measurements of stellar kinematics in the heart of the Milky Way and other local spirals, which will be soon available with LSST/Vera Rubin Observatory.

\begin{acknowledgments}
We thank the anonymous reviewer for providing feedback that greatly improved this work, and thank Stefano Carniani and Miroslava Dessauges-Zavadsky for fruitful discussions regarding the observability with JWST. We made use of pynbody (\url{https://github.com/pynbody/pynbody}) in our analysis for this paper. Simulations were performed on the Piz Daint supercomputer of the Swiss National Supercomputing Centre (CSCS) under the project id s1014. PRC, LM and TT acknowledge support from the Swiss National Science Foundation under the grant 200020\_178949. AB acknowledges support from the Natural Sciences and Engineering Research Council of Canada. PM acknowledges a NASA contract supporting the WFIRST-EXPO Science Investigation Team (15-WFIRST15-0004), administered by GSFC.\\
The data that support the findings of this study are available upon reasonable request from the authors.
\end{acknowledgments}

\bibliographystyle{aasjournal}
\bibliography{GigaEris}

\end{document}